\documentclass[twocolumn,prb,floatfix,nofootinbib,superscriptaddress]{revtex4-2}

\usepackage[utf8]{inputenc}
\usepackage[T1]{fontenc} 
\usepackage[english]{babel} 

\usepackage{amsmath}
\usepackage{multirow}
\usepackage{appendix}

\usepackage{amsthm}
\usepackage{amssymb}
\usepackage[normalem]{ulem}
\usepackage{listings}
\usepackage{xcolor}
\usepackage{cancel}
\definecolor{codegreen}{rgb}{0,0.6,0}
\definecolor{codegray}{rgb}{0.5,0.5,0.5}
\definecolor{codepurple}{rgb}{0.58,0,0.82}
\definecolor{backcolour}{rgb}{0.95,0.95,0.92}
\usepackage[colorlinks=true, citecolor=blue, linkcolor=blue, urlcolor=blue]{hyperref}

\lstdefinestyle{mystyle}{
    backgroundcolor=\color{backcolour},   
    commentstyle=\color{codegreen},
    keywordstyle=\color{magenta},
    numberstyle=\tiny\color{codegray},
    stringstyle=\color{codepurple},
    basicstyle=\ttfamily\footnotesize,
    breakatwhitespace=false,         
    breaklines=true,                 
    captionpos=b,                    
    keepspaces=true,                 
    numbersep=5pt,                  
    showspaces=false,                
    showstringspaces=false,
    showtabs=false,                  
    tabsize=1
}
\definecolor{dark-green}{RGB}{0, 128, 0}

\usepackage{enumitem}

\usepackage{dsfont}

\usepackage{graphicx}
\usepackage{color}

\usepackage{comment}
\usepackage{ stmaryrd }
\usepackage{physics}
\usepackage{tabularx}

\newenvironment{Aligned}
  {\begin{equation}
\begin{aligned}}
  {\end{aligned}
\end{equation}}

\begin{document}

\title{{Why finite-size non-Hermitian eigenvalues can miss asymptotic eigenstates in topological physics}}
\author{Lucien Jezequel}
\affiliation{Department of Physics, KTH Royal Institute of Technology, Stockholm 106 91, Sweden}
\author{Loïc Herviou}
\affiliation{Univ. Grenoble Alpes, CNRS, LPMMC, 38000 Grenoble, France}
\author{Jens H.~Bardarson}
\affiliation{Department of Physics, KTH Royal Institute of Technology, Stockholm 106 91, Sweden}

\begin{abstract}
    {Non-Hermitian systems exhibit a fundamental spectral dichotomy absent in Hermitian physics: the eigenvalue spectrum and the eigenstate spectrum can deviate significantly in the thermodynamic limit. }
    {We explain how extrapolating finite-size eigenvalues can miss important asymptotic eigenstates, with the unidirectional Hatano-Nelson model serving as a minimal realization of this phenomenon.}
    {Through exact analytical solutions, we show that this model contains not only hidden asymptotic modes but also quasi-Jordan structures that become macroscopic in the thermodynamic limit and appear more generally in systems with nontrivial bulk winding.}
    {Our framework explains how the apparent failure of the bulk-edge correspondence in models such as the non-Hermitian SSH chain can reflect the failure of finite-size eigenvalue extrapolation to detect asymptotic edge states in systems with a skin effect.}
    {These results clarify the limitations of using finite-size eigenvalues alone to infer thermodynamic-limit topology.}
\end{abstract}

\maketitle

{\it Introduction}---Non-Hermitian quantum mechanics has emerged as a powerful framework for describing open dissipative quantum systems, where effective Hamiltonians incorporate gain, loss \cite{feng2017non,liu2020gain}, and nonreciprocal couplings \cite{lau2018fundamental, Fruchart_2021,reisenbauer2024non}. 
While such descriptions are severe approximations for weakly dissipative systems \cite{Roccati_2022}, recent advances have revealed that non-Hermitian physics hosts genuinely new phenomena without Hermitian counterparts. 
Unlike Hermitian operators, whose eigenvalues are real and whose eigenstates form a complete orthogonal basis, non-Hermitian Hamiltonians exhibit a richer spectral structure, including complex eigenvalues \cite{Kawabata2019,wang2021topological,Schindler2023}, exceptional points \cite{Heiss_2004,Minganti2019,miri2019exceptional,Bergholtz2021,Wang2021,Liao2021,Delplace_2021}, and spectral singularities \cite{zhong2018winding,Ryu2022,Jezequel2023,Leclerc2024,Zhang2025}. 
These features enable diverse novel phenomena, such as enhanced sensing \cite{Wiersig2014,chen2017exceptional,hodaei2017enhanced,chen2019sensitivity,kononchuk2022exceptional,wiersig2020prospects}, unidirectional invisibility \cite{Longhi_2015,Brandstotter2019,Longhi_2022}, and non-Hermitian topological phases \cite{NonHermiPhysics,NonHemChernBands,TopoPhase,TopoBandNonHerm,Bergholtz2021}.

A prominent manifestation of the anomalous non-Hermitian spectral properties is the non-Hermitian skin effect, where all eigenstates become exponentially localized at system boundaries under open boundary conditions \cite{zhou2018observation,Yao2018, okuma2020topological,Borgnia2020,zhang2022review}. 
Although the effect was initially discovered in one-dimensional models, it is more universal and extends to both higher dimensions \cite{zhang2022universal,Wang2024,hu2025topological,Yang_2025} and interacting systems \cite{Shen_2022,Yang_2022, Alsallom2022}. 
In particular, with the skin effect there is an important difference between periodic and open boundary spectra: their band gaps may close at different parameter values, and a naive eigenvalue-based bulk-boundary correspondence appears to fail \cite{Lee_2016,xiong2018does,Fulga2024}.
Several approaches to reconcile these concepts have been discussed, such as the introduction of the generalized Brillouin zone \cite{Yao2018,Song_2019,Yang2020}, bi-orthogonal approaches \cite{Kunst18} or proposals to focus on the stability of singular values \cite{Herviou_2019,Porras2019,Brunelli2023,Monkman_2025}. 

In this work, we show that {one source of the apparent failure of the bulk-edge correspondence is the use of finite-size eigenvalue spectra to extrapolate the behavior of the thermodynamic limit.}
{For non-normal Hamiltonians, this extrapolation can fail because the finite-size eigenvalue spectrum and the asymptotic eigenstate spectrum may not coincide} \cite{Monkman_2025,mardani2025}.
{This is different from Hermitian physics, where approximate eigenstates remain uniformly tied to nearby eigenvalues uniformly in system size.}
Using the Hatano-Nelson model as a building block, and the non-Hermitian Su-Schrieffer-Heeger (SSH) chain as an illustration, we identify {asymptotic modes and quasi-Jordan structures that are not captured by the extrapolation of finite-size eigenvalues in many skin-effect systems.}
{This explains how finite-size eigenvalue-based diagnostics can fail to detect asymptotic topological edge states in models like the non-Hermitian SSH chain.}

{\it Formal definitions.}---The spectrum of a matrix or an operator $H$ can be defined in two different ways:
\begin{itemize}
    \item[] \textit{Eigenvalue Spectrum:} The set $\sigma_{\text{eig}}(H)$ of all eigenvalues of the operator $H$, which are the roots of the characteristic polynomial $\det(H-\lambda \mathds{1})$.
    
    \item[] \textit{Eigenstate Spectrum:} The set $\sigma_{\text{states}}(H)$ of energies $E$ for which there exists a state $\ket{\psi} $ such that $(H - E)\ket{\psi} = 0$.
\end{itemize}

For a finite matrix, these definitions coincide, $\sigma_{\text{eig}}(H) = \sigma_{\text{states}}(H)$: a nonzero solution of $(H-E)\ket{\psi}=0$ exists if and only if $E$ is an eigenvalue, including at degeneracies.
They also coincide for any Hermitian operator, finite or infinite, due to the stability relation
\begin{equation}
    \|(H - E)\ket{\psi}\| = \epsilon \implies \text{Dist}(E, \sigma_{\text{eig}}(H)) \leq \epsilon,
\end{equation}
where $\text{Dist}(E, \sigma_{\text{eig}}(H))$ is the distance of $E$ to the closest eigenvalue of the spectrum.
This follows from the spectral theorem: in the eigenbasis $\ket{\psi_k}$ where $ H\ket{\psi_k} = E_k\ket{\psi_k}$, with energies $E_k$, an arbitrary state $\ket{\psi}=\sum_k a_k\ket{\psi_k}$ satisfies
\begin{Aligned}
\|(H - E)\ket{\psi}\|^2 &= \sum_k |E_k - E|^2 |a_k|^2 \\&\geq \text{Dist}(E, \sigma_{\text{eig}}(H))^2 \|\psi\|^2,
\label{eq:2}\end{Aligned}
\hspace{-0.2cm}and therefore $\text{Dist}(E, \sigma_{\text{eig}}(H)) \leq \epsilon$. 
Importantly, in the above we used the orthonormality of the eigenbasis: $\langle\psi_k|\psi_{k'}\rangle=\delta_{k,k'}$, a property of Hermitian systems.
Conversely, if $E$ is an energy with $\text{Dist}(E, \sigma_{\text{eig}}(H)) = \epsilon$, then by definition there exists an eigenstate $\ket{\psi'}$ with energy $E'$ such that $|E-E'|=\epsilon$ and 
\begin{equation}
    \|(H - E)\ket{\psi'}\|=|E'-E|= \epsilon. \label{eq:3}
\end{equation}
{The set of energies $E$ for which there exists a state $\ket{\psi}$ such that $\|(H-E)\ket{\psi}\|\leq\epsilon\|\psi\|$ is called the $\epsilon$-\textit{pseudospectrum} \cite{trefethen2020spectra,flynn2023topological} of $H$; we denote this set $\sigma_\epsilon(H)$.}

For Hermitian operators, the constraints \eqref{eq:2} and \eqref{eq:3} ensure that the $\epsilon$-pseudospectrum always lies within a distance $\epsilon$ of the eigenvalue spectrum.
{This result is no longer true for non-Hermitian operators, where non-normality can make the pseudospectrum much larger than an $\epsilon$-neighborhood of the eigenvalue spectrum.}
{This has important consequences for extrapolating finite systems to the thermodynamic limit.}
Indeed, for a system of size $L$ with Hamiltonian $H_L$ there are in the thermodynamic limit $L \rightarrow \infty$ two different ways to generalize the notion of the spectrum. 
The eigenvalue spectrum can be defined as the limit of the roots of the characteristic polynomial, which is formally expressed as
\begin{equation}
    \sigma_\text{eig}(H_\infty) = \lim_{L\xrightarrow{}\infty}\lim_{\epsilon \xrightarrow{}0} \sigma_\epsilon(H_L).
\end{equation}
{Meanwhile, the eigenstate spectrum can be defined as the set of energies $E$ for which there exist normalized states $\ket{\psi_L}$ such that $\|(H_L-E)\ket{\psi_L}\|\xrightarrow{L \rightarrow \infty}0$, which is formally expressed as}
\begin{equation}
    \sigma_\text{states}(H_\infty) =\lim_{\epsilon \xrightarrow{}0}\lim_{L\xrightarrow{}\infty} \sigma_\epsilon(H_L).
\end{equation}
{The fact that these two limits may not coincide is a physically important feature of non-Hermitian operators with strong non-normality \cite{Okuma2021, flynn2023topological}.}
{It is the starting point for the discussion below.}

\begin{figure}[tb]
    \centering
    \includegraphics[width=1\linewidth]{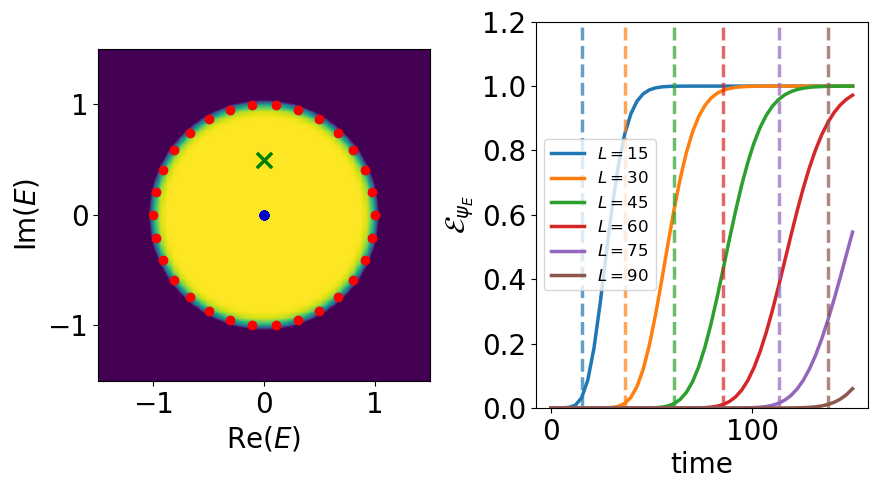}
    \caption{a) Bulk spectrum of the leftward-hopping Hatano-Nelson model \eqref{eq:Jordanblock} for $L=30$ sites, showing periodic (red) and open boundary (blue) eigenvalue spectra. 
    The yellow background denotes the $\epsilon$-pseudospectrum of the open system with $\epsilon=10^{-3}$.  
    {The green cross is an element of the pseudo-spectrum which is not part of the eigenvalue spectrum; it marks the value of $E$ used in panel b.}
    b) Temporal stability of the normalized state error \eqref{eq:error} for the almost eigenstate $\ket{\psi_E}$ in Eq.~\eqref{eq:JordanState}, evaluated at the green marker in panel a). The vertical lines indicate the time until which the error remains below 1\%, showing the growth of the lifetime with system size.
    }
    \label{fig:spectral_failure}
\end{figure}

{\it The Hatano-Nelson chain, a minimal model.}---{A simple model for which the thermodynamic eigenstate spectrum and the finite-size eigenvalue spectrum differ is the leftward-hopping Hatano-Nelson model} \cite{HatanoNelson,hatano1998non,TopoPhase,TopoBandNonHerm,Yao2018,Kawabata2019,Feng_2025} 
\begin{equation}
    H_L = \begin{pmatrix}
    0 & 1& & \\   
    & \ddots & \ddots & \\ 
    & & \ddots & 1 \\  
    & & & 0  
    \end{pmatrix}_{L \times L}, \label{eq:Jordanblock}
\end{equation}
with $L$ sites.
This model is both a physical system (a chain with hopping amplitudes) and a mathematical archetype: an order-$L$ Jordan block exceptional point.
{It is the minimal setting in which the order-of-limits distinction can be seen explicitly and tied to the dynamics of asymptotic states.}
While its eigenvalue spectrum is a single point ($\sigma_{\text{eig}}(H_L) = \{0\}$ for all $L$), it supports almost eigenstates for any $|E|<1$:
\begin{equation}
    \ket{\psi_E} = \sum_{n=0}^{L-1} E^n \ket{n}, \hspace{0.0cm} \|(H_L \hspace{-0.05cm}-\hspace{-0.05cm} E)\ket{\psi_E}\hspace{-0.07cm} \| = |E|^L \xrightarrow{L \to \infty} 0, \label{eq:JordanState}
\end{equation}
where $\ket{n}$ are the canonical basis vectors of \eqref{eq:Jordanblock}. 
In the thermodynamic (semi-infinite) limit $H_L \xrightarrow{L \xrightarrow{}\infty}H_\infty$, where the order of the exceptional point goes to infinity, $\sigma_{\text{states}}(H_\infty)$ therefore contains the entire unit disk $|E|<1$, which is much larger than the corresponding thermodynamic limit of the eigenvalue spectrum $\sigma_{\text{eig}}(H_L) \xrightarrow{L \xrightarrow{}\infty} \{0\}$ (see Fig \ref{fig:spectral_failure}a). 

In the thermodynamic limit ($L \to \infty$), as explicitly shown in Eq.~\eqref{eq:JordanState}, a sequence of states $|\psi_L\rangle$ may converge to an exact eigenstate of energy $E$ (satisfying $\|(H_L-E)\ket{\psi_L}\| \to 0$) while the energy $E$ may never come close to the eigenvalue spectrum $\sigma_{\text{eig}}(H_L)$ at any finite system size. 

{This phenomenon, where the thermodynamic limit supports asymptotic eigenstates at energies that remain far from the finite-size eigenvalue spectrum \cite{Herviou_2019,Monkman_2025,mardani2025}, illustrates why extrapolating finite-size eigenvalues can give an incomplete picture of large non-normal systems.}
{It is closely related to other familiar anomalous behavior of non-Hermitian matrices, like the extreme sensitivity to boundary conditions near high-order exceptional points.}
In the example of Eq.~\eqref{eq:JordanState}, a perturbation of order $\vert E \vert^L$ is sufficient to make $\ket{\psi_E}$ an actual eigenvector of the finite Hamiltonian. 
Equivalently, the eigenvalue spectrum $\sigma_{\text{eig}}(H_L)$, which without perturbation only contains $E=0$, can move throughout $|E|<1$ under exponentially small perturbations.
This issue typically appears in numerical algorithms where rounding errors due to the floating point precision play the role of a random perturbation \cite{trefethen2020spectra,PhysRevLett.125.226402,10.21468/SciPostPhys.17.5.128,Feng_2025}.
For instance, in the Hatano-Nelson model, a numerical diagonalization with 64-bit floats produces a spectrum that qualitatively deviates from the analytical result for even moderately sized chains with $L \approx 100$.
Evaluating the eigenvalue spectrum numerically is therefore, as is well known, inherently unstable for large systems close to an infinite order exceptional point. 

The existence of almost eigenstates, like $\ket{\psi_E}$ in \eqref{eq:JordanState}, has concrete physical implications.
To quantify the stability of the almost eigenstate we define the normalized state error 
\begin{equation}
\mathcal{E}_\psi(t) = \frac{\|e^{iHt}\vert\psi\rangle - e^{iEt}\vert \psi\rangle\|}{\|e^{iEt}\vert\psi\rangle\|},  \label{eq:error}
\end{equation}
which quantifies the temporal stability of an almost eigenstate $\vert \psi \rangle$ with complex energy $E$ under time evolution.
For exact eigenstates, it remains $0$ at all times.
For the states $\vert \psi_E \rangle$ in \eqref{eq:JordanState}, the time over which the approximation error remains small grows linearly with system size, as shown in Fig.~\ref{fig:spectral_failure}b.
This scaling provides a direct dynamical signature that these states behave as eigenstates on time scales that grow with system size and become exact eigenstates in the thermodynamic limit.

{\it Relationship between the eigenstate spectrum and alternative approaches to a non-Hermitian spectrum.}---
The eigenstate spectrum can be directly connected to two alternative notions of a spectrum in the literature.
The first is the singular value spectrum, derived from the decomposition $H = U S V^\dagger$, where $U$ and $V$ are unitary matrices and $S$ is a diagonal matrix of real non-negative singular values. 
This spectrum is stable under perturbations and has been shown to underpin conventional topological classifications in both Hermitian and non-Hermitian models \cite{Herviou_2019,TopoPhase,Kawabata2019,Liu2019}. 
The second is the notion of spectrum used in the mathematics literature, defined as the set of energies $E$ for which $(H-E)$ is not invertible, meaning that the Green's function $(H-E)^{-1}$ is not a bounded operator \cite{trefethen2020spectra,Henning2022}. 
These three approaches lead to equivalent definitions (see Appendix \ref{ap:B} for a proof) of the spectral gap $\Delta$ which are
\begin{align*}
\textit{Eigenstate criterion}:&\hspace{0.2cm}\forall \ket{\psi},  \|(H-E)\ket{\psi}\| \geq \Delta,  \\
\textit{Singular value criterion}:&\hspace{0.2cm}\sigma_{\mathrm{min}}[(H-E)^\dagger(H-E)] \geq \Delta^2,\\
\textit{Bounded inverse criterion}:&\hspace{0.2cm} \forall \ket{\psi}, \|(H-E)^{-1}\ket{\psi}\| \leq \Delta^{-1}.
\end{align*}
They therefore characterize the same spectrum defined as the set of energies $E$ where such gap $\Delta$ closes.

Here we only considered the action of $H$ on right states; similar definitions can be obtained for left states $\bra{\psi}$ by replacing $H-E$ by $(H-E)^\dagger$ in the above expressions. 
If there is a gap for both right and left states, the singular value criterion implies that the singular value spectrum of $H-E$ has a gap of size $\Delta$, which motivates the name of the criterion.

The fact that the eigenstate gap can be expressed as a Hermitian gap of $(H-E)^\dagger(H-E)$ or $(H-E)(H-E)^\dagger$ implies that it is stable under small perturbations in operator norm.
This should be contrasted with the different notion of an eigenvalue gap around an energy $E$, $\Delta'=\text{Dist}(E,\sigma_\text{eig}(H))$, which can be highly sensitive in non-normal systems.
Moreover, the eigenstate spectrum shares a key property with Hermitian systems: in the thermodynamic limit, the spectrum under open boundary conditions contains the bulk bands of the periodic boundary spectrum (see appendix~\ref{ap:A}). This is not generally true for non-Hermitian eigenvalue spectra because of the skin effect.
Therefore, the eigenstate spectrum offers a much stronger theoretical stability guarantee.
The only drawback of the eigenstate spectrum is computational: determining the eigenstate spectrum requires us to find the smallest singular value of $H-E$ for each energy $E$ of interest, while the eigenvalue spectrum only requires us to solve one linear algebra problem.
However the additional computational cost is compensated by the increased robustness of the eigenstate spectrum.

{\it Index theorem and quasi-Jordan structure.}---We next explain how the Jordan-chain structure visible in the Hatano-Nelson model is connected to point-gap topology more generally.
To this end, we start from the well-known index theorem that relates the winding number $W$ of a bulk non-Hermitian Hamiltonian to the number of right and left edge states at the boundary \cite{KaneLubensky,jezequel2024}.
Taking $H_\mathrm{bulk}$ to be the periodic Hamiltonian and $H$ the corresponding one with open boundary conditions, the index theorem states that as long as $H_\mathrm{bulk}$ has a (point) gap at $E$ we have
\begin{align}
            &\text{Ind}(H-E) \equiv \dim[\ker(H-E)] - \dim[\ker(H^\dagger-\Bar{E})] \notag \\
            &= W(H_\mathrm{\small bulk}-E)   \equiv \int dk \Tr\left[(H_\mathrm{\small bulk}-E)^{-1} \partial_kH_\mathrm{\small bulk}(k)\right].
\end{align}
We used $\Bar{E}$ to denote the complex conjugate of $E$.
This relation ensures the existence of topologically protected eigenstates when $W(H_\mathrm{bulk}-E)\neq 0$ (which is only possible for non-Hermitian operators). 
In fact, such nontrivial winding has additional consequences due to the identity \( W((H-E)^n) = nW(H-E) \) for any integer $n$ which means that $\text{Ind}[(H-E)^n]=n\text{Ind}[H-E]$. 
Therefore $(H-E)^n$ must have $n$-times more eigenstates than $H-E$. 
Such additional states are called generalized eigenstates of $H$ as they verify $(H-E)^n\ket{\psi}=0$ but not $(H-E)\ket{\psi}=0$ \cite{NonHermiPhysics}. 
This leads to an important implication: a nonzero winding number, \(W(H-E) \neq 0\), is associated with a proliferation of generalized eigenmodes.
Specifically, if \( \ket{\psi} \) is one of these new generalized eigenmodes satisfying \((H-E)^n\ket{\psi}=0\) but \((H-E)^{n-1}\ket{\psi}\neq0\), one can verify that the states \( \{(H-E)^k \ket{\psi}, 0\leq k\leq n-1\} \) form a basis on which \(H-E\) acts as a Jordan block of size \(n\), as in Eq.~\eqref{eq:Jordanblock}. 
This important result holds for any integer \(n\) as long as $H^n$ is sufficiently short-range compared to the lattice size so when $n\ll L$.
Consequently, when $H-E$ has nontrivial winding, the large open system supports a hierarchy of generalized modes and is asymptotically close to high-order Jordan-chain structures that can be hidden from the finite-size eigenvalue spectrum.
Note that this proliferation is directly related to the so-called non-Hermitian skin effect which states that a macroscopic number of boundary states exist as long as the index of the Hamiltonian is nontrivial \cite{Zhang_2020,okuma2020topological}.

We already discussed that the minimal model in Eq.~\eqref{eq:Jordanblock} has asymptotic eigenstates for $E\neq0$ that are not visible from the finite-size eigenvalue spectrum $\sigma_\text{eig}(H)=\{0\}$.
One can similarly construct asymptotic Jordan chains in the thermodynamic limit centered at such energies.
Indeed, one can verify that for $|E|<1$ the Hamiltonian \eqref{eq:Jordanblock} acts (up to exponentially small corrections) on the states
\begin{equation}
    \ket{\psi_{E;k}}=\sum_n P_k(n)E^{n-k}\ket{n}
\end{equation} 
where $P_k$ are the polynomials defined recursively by $P_0(n)=1$ and
\begin{equation}
    P_k(n+1)-P_k(n)=P_{k-1}(n),
\end{equation} 
and $k=1,\ldots,m$, as a single macroscopic Jordan block of order $m$. 
This approximate Jordan-block structure holds as long as the states $\ket{\psi_{E;1,\dots,m}}$ have a negligible amplitude at the right boundary, which is satisfied whenever $m \ll L$.
These new Jordan blocks are, by definition, exceptional points and this therefore illustrates how high-order exceptional points emerge generically in any non-Hermitian system with topological winding, while often being hidden from the finite size eigenvalue spectrum $\sigma_\text{eig}(H)$.

\begin{figure}[t]
    \centering
    \includegraphics[width=0.52\linewidth]{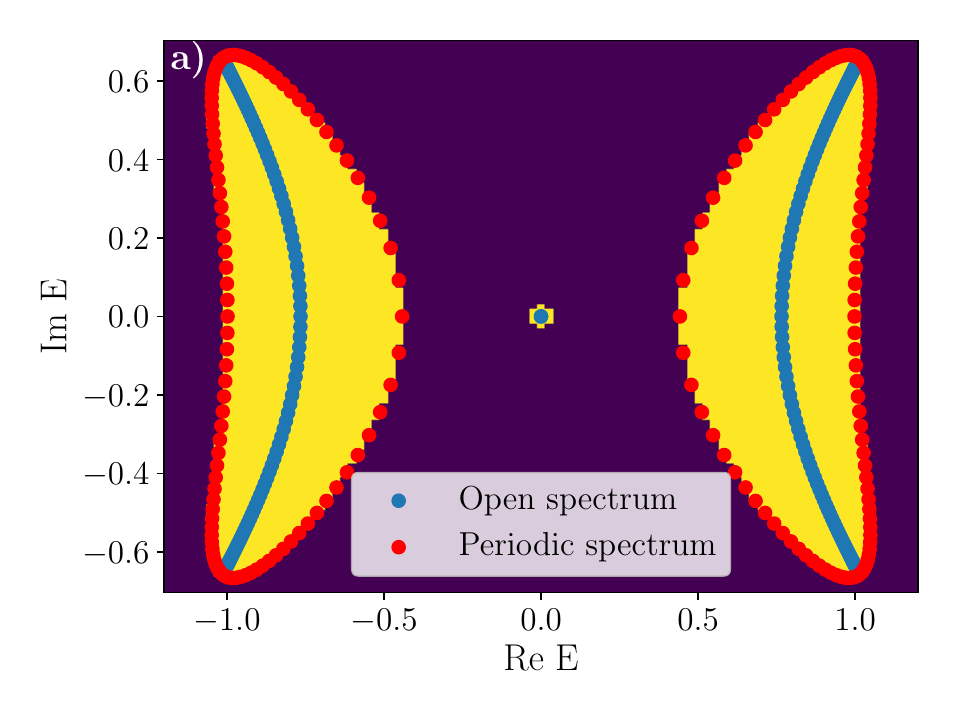} \hspace{-0.54cm}
    \includegraphics[width=0.52\linewidth]{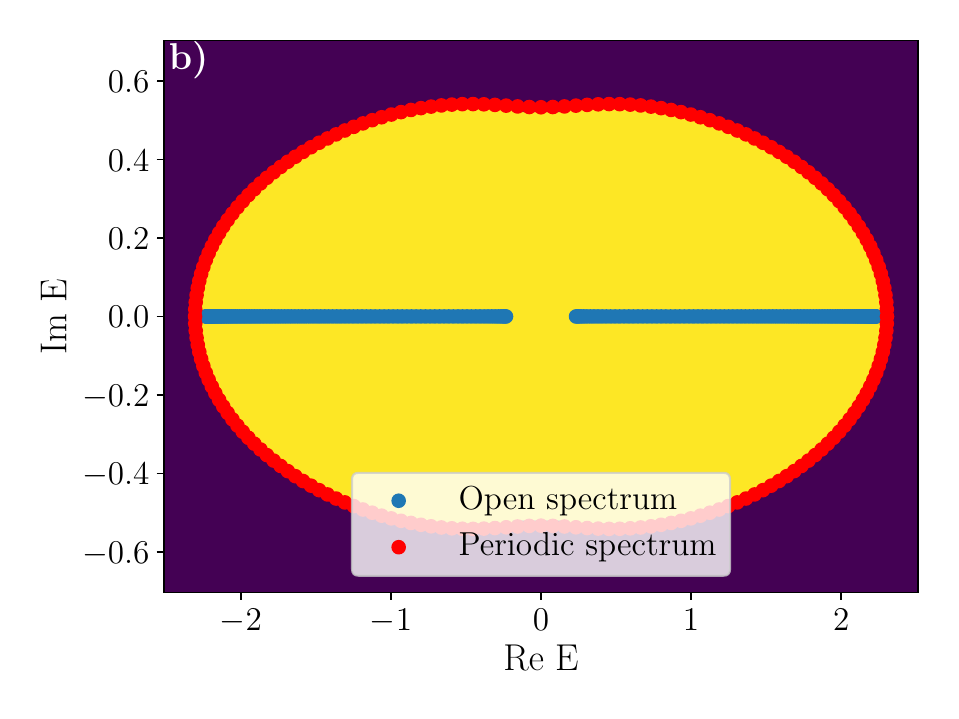}\\
    \vspace{-0.2cm}\includegraphics[width=1\linewidth]{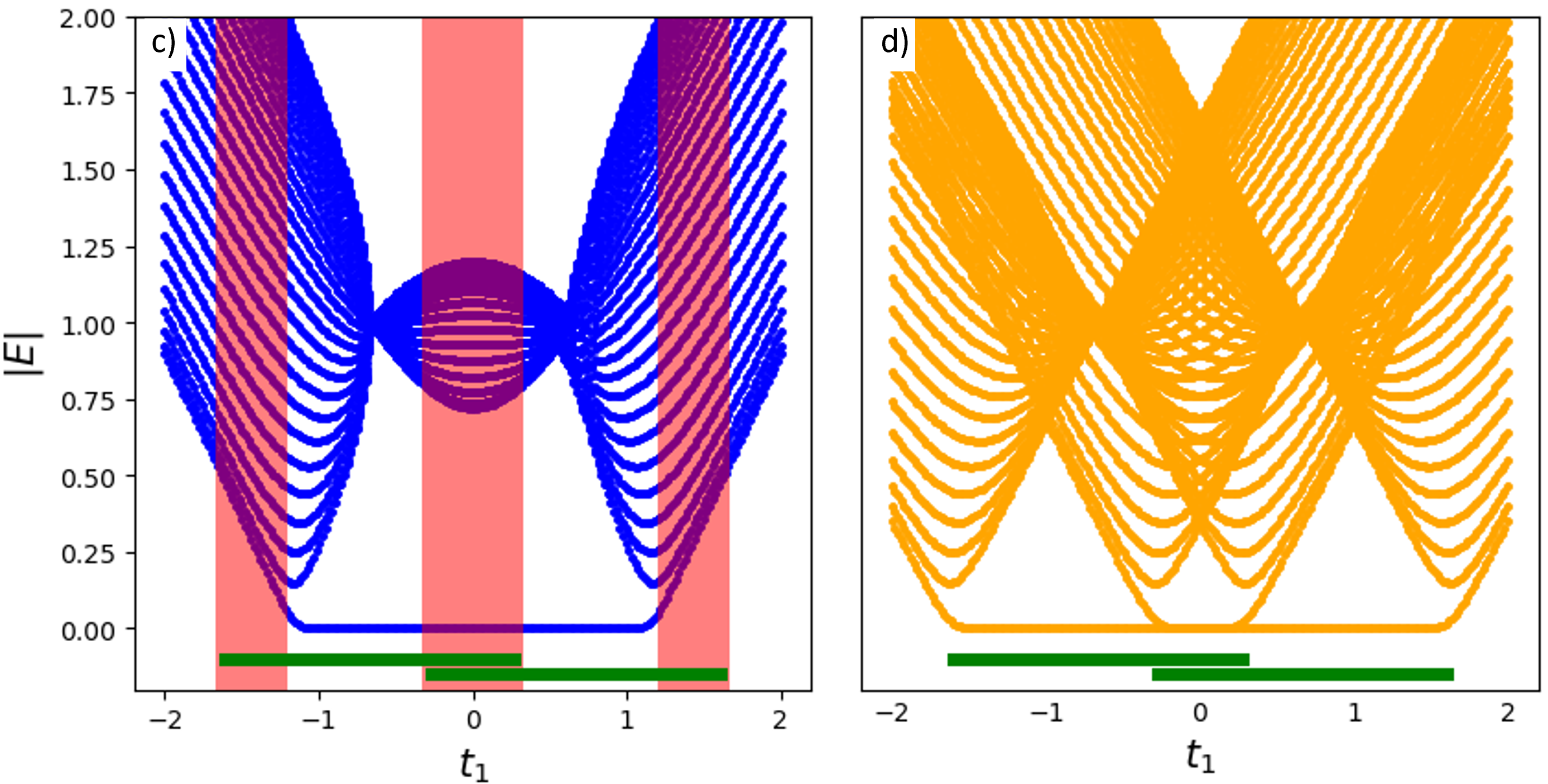}
    \caption{a) Spectrum of the non-Hermitian SSH model defined in Eq.~\eqref{eq:NHSSHchain} for $(t_1, t_2,\gamma)=(0.2, 1, 4/3)$ and $L = 100$. The dots correspond to the eigenvalue spectrum for periodic and open boundary conditions, while the yellow background is the eigenstate spectrum of the open system. It is the full area enclosed by the periodic spectrum. b) Similar spectrum for $(t_1, t_2,\gamma)=(1.4, 1, 4/3)$. Note that now eigenstates with zero energy are present despite the absence of topological modes in the eigenvalue spectrum. 
    Plot of c) eigenvalue spectrum  and d) singular value spectrum for $(t_2,\gamma)=(1,4/3)$ and  $L=20$ sites with open boundary conditions, reproducing the configuration of \cite{Yao2018}.
    Parameters for which topological edge modes at zero energy exist are denoted in green. The eigenvalue spectrum is deficient in the red regions with either a gap while there is one mode at zero energy ($1.2\lesssim|t_1|\leq 5/3$) or a single zero eigenvalue where there are two modes at zero energy ($|t_1|\leq 1/3$). 
    {The singular-value spectrum therefore detects these zero-energy modes in the regions where the eigenvalue criterion alone gives an incomplete diagnosis.}
    }
    \label{fig:NHSSHchain}
\end{figure}

{\it A case study: the non-Hermitian SSH model.}---As a more complex example of eigenstates and exceptional points being hidden from the finite size eigenvalue spectrum, we study the non-Hermitian SSH model and reinterpret its anomalous behavior.
This model is widely discussed in the non-Hermitian literature \cite{Yao2018,Yin2018,Herviou_2019,Song_2019,Bergholtz2021,lin_topological_2023} and its Hamiltonian is given by
Its Hamiltonian is given by
\begin{Aligned}
    H_\mathrm{nH-SSH} &= \hspace{-0.1cm}\sum_n (t_1-\gamma/2)\hspace{-0.05cm}\ket{n,2}\hspace{-0.1cm} \bra{n,1} \hspace{-0.05cm}+ \hspace{-0.05cm}(t_1+\gamma/2)\hspace{-0.05cm}\ket{n,1} \hspace{-0.1cm}\bra{n,2} \\
    &+ t_2(\ket{n+1, 1} \hspace{-0.1cm}\bra{n,2}\hspace{-0.05cm} + \hspace{-0.05cm}\ket{n,2} \hspace{-0.1cm}\bra{n+1,1}),\label{eq:NHSSHchain}
\end{Aligned}
\noindent \hspace{-0.3cm} where $\ket{n, \alpha}$ describes the orbital $\alpha$ at position $n$ and
all coefficients are taken positive for simplicity.
The eigenvalues $E_k$ of the bulk periodic Hamiltonian at momentum $k$ satisfy
\begin{equation}
     E_k^2 = \left(t_1 + t_2 \cos k \right)^2 - \left(\frac{\gamma}{2} - i t_2 \sin k \right)^2,
\end{equation}
and the standard bulk gap closes for $\gamma^2 = 4 (t_1 \pm t_2)^2$.
With open boundary conditions, the eigenvalue phase diagram appears significantly different, with an apparent transition at $\gamma^2 = 4(t_1^2 \pm t_2^2)$ with an intermediate phase hosting topological modes at zero energy (the gap closing point is obtained analytically using a non-unitary gauge transformation that absorbs the difference between left and right hopping) \cite{Yao2018,lin_topological_2023}.
In fact, the eigenstate spectrum of the model with open boundary conditions differs significantly from its eigenvalue spectrum (see Fig.~\ref{fig:NHSSHchain}a and Fig.~\ref{fig:NHSSHchain}b).
For simplicity, we focus on the zero-energy eigenstates even if a similar analysis could be done at any energy.
We place ourselves in the semi-infinite limit.
Let $a_n$ and $b_n$ be the wave-function amplitudes on the $\ket{n,1}$ and $\ket{n,2}$ sites, respectively. 
The modes at zero energy are known~\cite{Yao2018,Kunst18} to exhibit exponential localization at the left edge with   
\begin{equation}
     a_n=\left(\frac{-t_1+\gamma/2}{t_2}\right)^n, b_n=0,
    \label{eq:Leftedge}
\end{equation}
when $|-t_1+\gamma/2|<|t_2|$, and at the right edge with 
\begin{equation}
 a_n=0, b_n=\left(-\frac{t_2}{t_1+\gamma/2}\right)^{n-L}, 
    \label{eq:Rightedge}
\end{equation}
when $|t_1+\gamma/2|<|t_2|$.
For finite systems, these become exponentially precise approximate modes at zero energy with errors of order $(\frac{-t_1\pm\gamma/2}{t_2})^L$. 
This is different from conventional eigenvalue analysis, as there is a vanishing eigenvalue $E\xrightarrow{L\xrightarrow{}\infty}0$ only when $4(t_1^2 - t_2^2) \leq \gamma^2 \leq 4(t_1^2 + t_2^2)$. 
This is different from the condition $|t_1\pm\gamma/2|\leq t_2$ for the existence of an exponentially precise eigenstate (see the red regions in Fig.~\ref{fig:NHSSHchain}(c)). 
Therefore we see that the closing of the bulk gap for values $|t_1\pm\gamma/2|= t_2$ corresponds to the appearance or disappearance of an exponentially precise edge eigenstate but not to the appearance or disappearance of a vanishing eigenvalue.
{This discrepancy between the closing of the bulk gap and the point where an asymptotic eigenstate appears at $E=0$ for open boundary conditions, is one standard manifestation of the modified bulk-boundary correspondence in non-Hermitian systems \cite{Lee_2016,Yao2018}.}
{Such discrepancy can be interpreted as a regime where extrapolating the finite-size eigenvalue spectrum does not faithfully track the asymptotic zero-energy edge states, while the singular-value/eigenstate spectrum does.}
{This is the same order-of-limits phenomenon already visible in the simpler Hatano-Nelson model, now appearing in a topological setting where it changes the thermodynamic edge-state diagnosis.}

Finally, we show that the non-Hermitian SSH model, like the Hatano-Nelson model, hosts an asymptotic family of states $\ket{\psi_k}$ forming a quasi-Jordan chain with $H \ket{\psi_k}= \ket{\psi_{k-1}}$ up to finite-size corrections.
The coefficients $a_n^{(k)}$ and $b_n^{(k)}$ of those states take the form
\begin{Aligned}
    &a_n^{(k)}=P_k(n)r^n \hspace{0.3cm}b_n=0 &\text{ if $k$ odd}\\
    &a_n^{(k)}=0 \hspace{1.37cm}b_n=P_k(n) r^n &\text{ if $k$ even}
\end{Aligned}
\hspace{-0.18cm}with $r=\frac{-t_1+\gamma/2}{t_2}$ and $r'=\frac{-t_2}{t_1+\gamma/2}$ and $P_k(n)$ are polynomials generated recursively with $P_0(n)=1$ and the recursive relation
\begin{Aligned}
    &P_{k+1}(n+1)-r'/rP_{k+1}(n)=P_{k}(n)&\text{ if $k$ odd},\\
    &P_{k+1}(n+1)-P_{k+1}(n)=P_{k}(n)&\text{ if $k$ even}.
\end{Aligned}
\hspace{-0.33cm} For finite systems, the block survives up to exponential corrections as long as the modes have small amplitudes on the right edge which is true as long as $n \ll L$. 
This example illustrates how open non-Hermitian systems with skin effect can be close to macroscopic-order Jordan structures, despite having an apparently well-defined finite-size eigenvalue spectrum.

{\it Conclusions and discussions.}---{In this work, we have shown that some of the discrepancies arising in the study of strongly non-normal non-Hermitian systems originate from using finite-size eigenvalue spectra to extrapolate the existence of asymptotic eigenstates in the large-size limit.}
{The eigenstate spectrum is stable and well-defined through its relation to pseudospectra, singular values, and bounded Green's functions, and it gives a direct way to characterize this thermodynamic-limit behavior.}

{This mechanism is explicit in the Hatano-Nelson chain: finite-size eigenvalues collapse to a single point, while asymptotic eigenstates fill the disk $|E|\leq 1$ and behave dynamically as eigenstates for times that grow with system size.}
{We have further shown that the same instability can be understood through the proximity to quasi-Jordan structures that become macroscopic in the thermodynamic limit.}
{This shows that exceptional-point physics is not restricted to isolated finite-size degeneracies, rather it appears as an asymptotic organizing structure over broad spectral regions in all systems with nontrivial winding.}
{This leads to hidden asymptotic eigenstates with physical consequences: during time evolution they behave as eigenstates for macroscopic times before finite-size effects become important.}

{The non-Hermitian SSH model illustrates how the distinction between eigenvalue and eigenstate spectra affects the topology.}
{The appearance or disappearance of asymptotic zero-energy edge states is detected by the eigenstate or singular-value spectrum, even in regimes where extrapolating the finite-size eigenvalues alone can give an incomplete picture of the dynamic of large non-Hermitian systems.}
{Thus, while eigenvalue methods need not be abandoned, their finite-size extrapolation must be treated with care in skin-effect systems.}
Extensions to weakly interacting bosonic systems and to fermionic many-body settings, where Pauli effects can produce additional localization phenomena \citep{PhysRevLett.127.070402,Alsallom2022, Zheng2024, Hamanaka2025, Zhong2025}, are natural directions for future work.

\paragraph{Acknowledgment} We thank Emil Bergholtz and Fan Yang for insightful discussions.  L.J. and J.H.B. acknowledge financial support by the Swedish Research Council (VR) through Grant No. 2020-00214, and the European Research Council (ERC) under the European Union’s Horizon 2020 research and innovation program (Grant Agreement No. 101001902). L.H. acknowledges financial support by the CNRS through the Tremplin funding.

\bibliographystyle{apsrev4-2}
\bibliography{bibliography}

\begin{thebibliography}{76}%
\makeatletter
\providecommand \@ifxundefined [1]{%
 \@ifx{#1\undefined}
}%
\providecommand \@ifnum [1]{%
 \ifnum #1\expandafter \@firstoftwo
 \else \expandafter \@secondoftwo
 \fi
}%
\providecommand \@ifx [1]{%
 \ifx #1\expandafter \@firstoftwo
 \else \expandafter \@secondoftwo
 \fi
}%
\providecommand \natexlab [1]{#1}%
\providecommand \enquote  [1]{``#1''}%
\providecommand \bibnamefont  [1]{#1}%
\providecommand \bibfnamefont [1]{#1}%
\providecommand \citenamefont [1]{#1}%
\providecommand \href@noop [0]{\@secondoftwo}%
\providecommand \href [0]{\begingroup \@sanitize@url \@href}%
\providecommand \@href[1]{\@@startlink{#1}\@@href}%
\providecommand \@@href[1]{\endgroup#1\@@endlink}%
\providecommand \@sanitize@url [0]{\catcode `\\12\catcode `\$12\catcode `\&12\catcode `\#12\catcode `\^12\catcode `\_12\catcode `\%12\relax}%
\providecommand \@@startlink[1]{}%
\providecommand \@@endlink[0]{}%
\providecommand \url  [0]{\begingroup\@sanitize@url \@url }%
\providecommand \@url [1]{\endgroup\@href {#1}{\urlprefix }}%
\providecommand \urlprefix  [0]{URL }%
\providecommand \Eprint [0]{\href }%
\providecommand \doibase [0]{https://doi.org/}%
\providecommand \selectlanguage [0]{\@gobble}%
\providecommand \bibinfo  [0]{\@secondoftwo}%
\providecommand \bibfield  [0]{\@secondoftwo}%
\providecommand \translation [1]{[#1]}%
\providecommand \BibitemOpen [0]{}%
\providecommand \bibitemStop [0]{}%
\providecommand \bibitemNoStop [0]{.\EOS\space}%
\providecommand \EOS [0]{\spacefactor3000\relax}%
\providecommand \BibitemShut  [1]{\csname bibitem#1\endcsname}%
\let\auto@bib@innerbib\@empty
\bibitem [{\citenamefont {Feng}\ \emph {et~al.}(2017)\citenamefont {Feng}, \citenamefont {El-Ganainy},\ and\ \citenamefont {Ge}}]{feng2017non}%
  \BibitemOpen
  \bibfield  {author} {\bibinfo {author} {\bibfnamefont {L.}~\bibnamefont {Feng}}, \bibinfo {author} {\bibfnamefont {R.}~\bibnamefont {El-Ganainy}},\ and\ \bibinfo {author} {\bibfnamefont {L.}~\bibnamefont {Ge}},\ }\href {https://doi.org/10.1038/s41566-017-0031-1} {\bibfield  {journal} {\bibinfo  {journal} {Nature Photonics}\ }\textbf {\bibinfo {volume} {11}},\ \bibinfo {pages} {752} (\bibinfo {year} {2017})}\BibitemShut {NoStop}%
\bibitem [{\citenamefont {Liu}\ \emph {et~al.}(2020)\citenamefont {Liu}, \citenamefont {Ma}, \citenamefont {Yang}, \citenamefont {Zhang}, \citenamefont {Gao}, \citenamefont {Xiang}, \citenamefont {Cui},\ and\ \citenamefont {Zhang}}]{liu2020gain}%
  \BibitemOpen
  \bibfield  {author} {\bibinfo {author} {\bibfnamefont {S.}~\bibnamefont {Liu}}, \bibinfo {author} {\bibfnamefont {S.}~\bibnamefont {Ma}}, \bibinfo {author} {\bibfnamefont {C.}~\bibnamefont {Yang}}, \bibinfo {author} {\bibfnamefont {L.}~\bibnamefont {Zhang}}, \bibinfo {author} {\bibfnamefont {W.}~\bibnamefont {Gao}}, \bibinfo {author} {\bibfnamefont {Y.~J.}\ \bibnamefont {Xiang}}, \bibinfo {author} {\bibfnamefont {T.~J.}\ \bibnamefont {Cui}},\ and\ \bibinfo {author} {\bibfnamefont {S.}~\bibnamefont {Zhang}},\ }\href {https://doi.org/10.1103/PhysRevApplied.13.014047} {\bibfield  {journal} {\bibinfo  {journal} {Phys. Rev. Appl.}\ }\textbf {\bibinfo {volume} {13}},\ \bibinfo {pages} {014047} (\bibinfo {year} {2020})}\BibitemShut {NoStop}%
\bibitem [{\citenamefont {Lau}\ and\ \citenamefont {Clerk}(2018)}]{lau2018fundamental}%
  \BibitemOpen
  \bibfield  {author} {\bibinfo {author} {\bibfnamefont {H.-K.}\ \bibnamefont {Lau}}\ and\ \bibinfo {author} {\bibfnamefont {A.~A.}\ \bibnamefont {Clerk}},\ }\href {https://doi.org/10.1038/s41467-018-06477-7} {\bibfield  {journal} {\bibinfo  {journal} {Nature Communications}\ }\textbf {\bibinfo {volume} {9}},\ \bibinfo {pages} {4320} (\bibinfo {year} {2018})}\BibitemShut {NoStop}%
\bibitem [{\citenamefont {Fruchart}\ \emph {et~al.}(2021)\citenamefont {Fruchart}, \citenamefont {Hanai}, \citenamefont {Littlewood},\ and\ \citenamefont {Vitelli}}]{Fruchart_2021}%
  \BibitemOpen
  \bibfield  {author} {\bibinfo {author} {\bibfnamefont {M.}~\bibnamefont {Fruchart}}, \bibinfo {author} {\bibfnamefont {R.}~\bibnamefont {Hanai}}, \bibinfo {author} {\bibfnamefont {P.~B.}\ \bibnamefont {Littlewood}},\ and\ \bibinfo {author} {\bibfnamefont {V.}~\bibnamefont {Vitelli}},\ }\href {https://doi.org/10.1038/s41586-021-03375-9} {\bibfield  {journal} {\bibinfo  {journal} {Nature}\ }\textbf {\bibinfo {volume} {592}},\ \bibinfo {pages} {363–369} (\bibinfo {year} {2021})}\BibitemShut {NoStop}%
\bibitem [{\citenamefont {Reisenbauer}\ \emph {et~al.}(2024)\citenamefont {Reisenbauer}, \citenamefont {Rudolph}, \citenamefont {Egyed}, \citenamefont {Hornberger}, \citenamefont {Zasedatelev}, \citenamefont {Abuzarli}, \citenamefont {Stickler},\ and\ \citenamefont {Delić}}]{reisenbauer2024non}%
  \BibitemOpen
  \bibfield  {author} {\bibinfo {author} {\bibfnamefont {M.}~\bibnamefont {Reisenbauer}}, \bibinfo {author} {\bibfnamefont {H.}~\bibnamefont {Rudolph}}, \bibinfo {author} {\bibfnamefont {L.}~\bibnamefont {Egyed}}, \bibinfo {author} {\bibfnamefont {K.}~\bibnamefont {Hornberger}}, \bibinfo {author} {\bibfnamefont {A.~V.}\ \bibnamefont {Zasedatelev}}, \bibinfo {author} {\bibfnamefont {M.}~\bibnamefont {Abuzarli}}, \bibinfo {author} {\bibfnamefont {B.~A.}\ \bibnamefont {Stickler}},\ and\ \bibinfo {author} {\bibfnamefont {U.}~\bibnamefont {Delić}},\ }\href {https://doi.org/10.1038/s41567-024-02589-8} {\bibfield  {journal} {\bibinfo  {journal} {Nature Physics}\ }\textbf {\bibinfo {volume} {20}},\ \bibinfo {pages} {1629} (\bibinfo {year} {2024})}\BibitemShut {NoStop}%
\bibitem [{\citenamefont {Roccati}\ \emph {et~al.}(2022)\citenamefont {Roccati}, \citenamefont {Palma}, \citenamefont {Ciccarello},\ and\ \citenamefont {Bagarello}}]{Roccati_2022}%
  \BibitemOpen
  \bibfield  {author} {\bibinfo {author} {\bibfnamefont {F.}~\bibnamefont {Roccati}}, \bibinfo {author} {\bibfnamefont {G.~M.}\ \bibnamefont {Palma}}, \bibinfo {author} {\bibfnamefont {F.}~\bibnamefont {Ciccarello}},\ and\ \bibinfo {author} {\bibfnamefont {F.}~\bibnamefont {Bagarello}},\ }\href {https://doi.org/10.1142/S1230161222500044} {\bibfield  {journal} {\bibinfo  {journal} {Open Systems \& Information Dynamics}\ }\textbf {\bibinfo {volume} {29}},\ \bibinfo {pages} {2250004} (\bibinfo {year} {2022})}\BibitemShut {NoStop}%
\bibitem [{\citenamefont {Kawabata}\ \emph {et~al.}(2019)\citenamefont {Kawabata}, \citenamefont {Shiozaki}, \citenamefont {Ueda},\ and\ \citenamefont {Sato}}]{Kawabata2019}%
  \BibitemOpen
  \bibfield  {author} {\bibinfo {author} {\bibfnamefont {K.}~\bibnamefont {Kawabata}}, \bibinfo {author} {\bibfnamefont {K.}~\bibnamefont {Shiozaki}}, \bibinfo {author} {\bibfnamefont {M.}~\bibnamefont {Ueda}},\ and\ \bibinfo {author} {\bibfnamefont {M.}~\bibnamefont {Sato}},\ }\href {https://doi.org/10.1103/PhysRevX.9.041015} {\bibfield  {journal} {\bibinfo  {journal} {Phys. Rev. X}\ }\textbf {\bibinfo {volume} {9}},\ \bibinfo {pages} {041015} (\bibinfo {year} {2019})}\BibitemShut {NoStop}%
\bibitem [{\citenamefont {Wang}\ \emph {et~al.}(2021{\natexlab{a}})\citenamefont {Wang}, \citenamefont {Dutt}, \citenamefont {Wojcik},\ and\ \citenamefont {Fan}}]{wang2021topological}%
  \BibitemOpen
  \bibfield  {author} {\bibinfo {author} {\bibfnamefont {K.}~\bibnamefont {Wang}}, \bibinfo {author} {\bibfnamefont {A.}~\bibnamefont {Dutt}}, \bibinfo {author} {\bibfnamefont {C.~C.}\ \bibnamefont {Wojcik}},\ and\ \bibinfo {author} {\bibfnamefont {S.}~\bibnamefont {Fan}},\ }\href {https://doi.org/10.1038/s41586-021-03848-x} {\bibfield  {journal} {\bibinfo  {journal} {Nature}\ }\textbf {\bibinfo {volume} {598}},\ \bibinfo {pages} {59} (\bibinfo {year} {2021}{\natexlab{a}})}\BibitemShut {NoStop}%
\bibitem [{\citenamefont {Schindler}\ \emph {et~al.}(2023)\citenamefont {Schindler}, \citenamefont {Gu}, \citenamefont {Lian},\ and\ \citenamefont {Kawabata}}]{Schindler2023}%
  \BibitemOpen
  \bibfield  {author} {\bibinfo {author} {\bibfnamefont {F.}~\bibnamefont {Schindler}}, \bibinfo {author} {\bibfnamefont {K.}~\bibnamefont {Gu}}, \bibinfo {author} {\bibfnamefont {B.}~\bibnamefont {Lian}},\ and\ \bibinfo {author} {\bibfnamefont {K.}~\bibnamefont {Kawabata}},\ }\href {https://doi.org/10.1103/PRXQuantum.4.030315} {\bibfield  {journal} {\bibinfo  {journal} {PRX Quantum}\ }\textbf {\bibinfo {volume} {4}},\ \bibinfo {pages} {030315} (\bibinfo {year} {2023})}\BibitemShut {NoStop}%
\bibitem [{\citenamefont {Heiss}(2004)}]{Heiss_2004}%
  \BibitemOpen
  \bibfield  {author} {\bibinfo {author} {\bibfnamefont {W.~D.}\ \bibnamefont {Heiss}},\ }\href {https://doi.org/10.1088/0305-4470/37/6/034} {\bibfield  {journal} {\bibinfo  {journal} {Journal of Physics A: Mathematical and General}\ }\textbf {\bibinfo {volume} {37}},\ \bibinfo {pages} {2455–2464} (\bibinfo {year} {2004})}\BibitemShut {NoStop}%
\bibitem [{\citenamefont {Minganti}\ \emph {et~al.}(2019)\citenamefont {Minganti}, \citenamefont {Miranowicz}, \citenamefont {Chhajlany},\ and\ \citenamefont {Nori}}]{Minganti2019}%
  \BibitemOpen
  \bibfield  {author} {\bibinfo {author} {\bibfnamefont {F.}~\bibnamefont {Minganti}}, \bibinfo {author} {\bibfnamefont {A.}~\bibnamefont {Miranowicz}}, \bibinfo {author} {\bibfnamefont {R.~W.}\ \bibnamefont {Chhajlany}},\ and\ \bibinfo {author} {\bibfnamefont {F.}~\bibnamefont {Nori}},\ }\href {https://doi.org/10.1103/PhysRevA.100.062131} {\bibfield  {journal} {\bibinfo  {journal} {Phys. Rev. A}\ }\textbf {\bibinfo {volume} {100}},\ \bibinfo {pages} {062131} (\bibinfo {year} {2019})}\BibitemShut {NoStop}%
\bibitem [{\citenamefont {Miri}\ and\ \citenamefont {Alù}(2019)}]{miri2019exceptional}%
  \BibitemOpen
  \bibfield  {author} {\bibinfo {author} {\bibfnamefont {M.-A.}\ \bibnamefont {Miri}}\ and\ \bibinfo {author} {\bibfnamefont {A.}~\bibnamefont {Alù}},\ }\href {https://doi.org/10.1126/science.aar7709} {\bibfield  {journal} {\bibinfo  {journal} {Science}\ }\textbf {\bibinfo {volume} {363}},\ \bibinfo {pages} {eaar7709} (\bibinfo {year} {2019})}\BibitemShut {NoStop}%
\bibitem [{\citenamefont {Bergholtz}\ \emph {et~al.}(2021)\citenamefont {Bergholtz}, \citenamefont {Budich},\ and\ \citenamefont {Kunst}}]{Bergholtz2021}%
  \BibitemOpen
  \bibfield  {author} {\bibinfo {author} {\bibfnamefont {E.~J.}\ \bibnamefont {Bergholtz}}, \bibinfo {author} {\bibfnamefont {J.~C.}\ \bibnamefont {Budich}},\ and\ \bibinfo {author} {\bibfnamefont {F.~K.}\ \bibnamefont {Kunst}},\ }\href {https://doi.org/10.1103/RevModPhys.93.015005} {\bibfield  {journal} {\bibinfo  {journal} {Rev. Mod. Phys.}\ }\textbf {\bibinfo {volume} {93}},\ \bibinfo {pages} {015005} (\bibinfo {year} {2021})}\BibitemShut {NoStop}%
\bibitem [{\citenamefont {Wang}\ \emph {et~al.}(2021{\natexlab{b}})\citenamefont {Wang}, \citenamefont {Xiao}, \citenamefont {Budich}, \citenamefont {Yi},\ and\ \citenamefont {Xue}}]{Wang2021}%
  \BibitemOpen
  \bibfield  {author} {\bibinfo {author} {\bibfnamefont {K.}~\bibnamefont {Wang}}, \bibinfo {author} {\bibfnamefont {L.}~\bibnamefont {Xiao}}, \bibinfo {author} {\bibfnamefont {J.~C.}\ \bibnamefont {Budich}}, \bibinfo {author} {\bibfnamefont {W.}~\bibnamefont {Yi}},\ and\ \bibinfo {author} {\bibfnamefont {P.}~\bibnamefont {Xue}},\ }\href {https://doi.org/10.1103/PhysRevLett.127.026404} {\bibfield  {journal} {\bibinfo  {journal} {Phys. Rev. Lett.}\ }\textbf {\bibinfo {volume} {127}},\ \bibinfo {pages} {026404} (\bibinfo {year} {2021}{\natexlab{b}})}\BibitemShut {NoStop}%
\bibitem [{\citenamefont {Liao}\ \emph {et~al.}(2021)\citenamefont {Liao}, \citenamefont {Leblanc}, \citenamefont {Ren}, \citenamefont {Li}, \citenamefont {Li}, \citenamefont {Solnyshkov}, \citenamefont {Malpuech}, \citenamefont {Yao},\ and\ \citenamefont {Fu}}]{Liao2021}%
  \BibitemOpen
  \bibfield  {author} {\bibinfo {author} {\bibfnamefont {Q.}~\bibnamefont {Liao}}, \bibinfo {author} {\bibfnamefont {C.}~\bibnamefont {Leblanc}}, \bibinfo {author} {\bibfnamefont {J.}~\bibnamefont {Ren}}, \bibinfo {author} {\bibfnamefont {F.}~\bibnamefont {Li}}, \bibinfo {author} {\bibfnamefont {Y.}~\bibnamefont {Li}}, \bibinfo {author} {\bibfnamefont {D.}~\bibnamefont {Solnyshkov}}, \bibinfo {author} {\bibfnamefont {G.}~\bibnamefont {Malpuech}}, \bibinfo {author} {\bibfnamefont {J.}~\bibnamefont {Yao}},\ and\ \bibinfo {author} {\bibfnamefont {H.}~\bibnamefont {Fu}},\ }\href {https://doi.org/10.1103/PhysRevLett.127.107402} {\bibfield  {journal} {\bibinfo  {journal} {Phys. Rev. Lett.}\ }\textbf {\bibinfo {volume} {127}},\ \bibinfo {pages} {107402} (\bibinfo {year} {2021})}\BibitemShut {NoStop}%
\bibitem [{\citenamefont {Delplace}\ \emph {et~al.}(2021)\citenamefont {Delplace}, \citenamefont {Yoshida},\ and\ \citenamefont {Hatsugai}}]{Delplace_2021}%
  \BibitemOpen
  \bibfield  {author} {\bibinfo {author} {\bibfnamefont {P.}~\bibnamefont {Delplace}}, \bibinfo {author} {\bibfnamefont {T.}~\bibnamefont {Yoshida}},\ and\ \bibinfo {author} {\bibfnamefont {Y.}~\bibnamefont {Hatsugai}},\ }\href {https://doi.org/10.1103/PhysRevLett.127.186602} {\bibfield  {journal} {\bibinfo  {journal} {Phys. Rev. Lett.}\ }\textbf {\bibinfo {volume} {127}},\ \bibinfo {pages} {186602} (\bibinfo {year} {2021})}\BibitemShut {NoStop}%
\bibitem [{\citenamefont {Zhong}\ \emph {et~al.}(2018)\citenamefont {Zhong}, \citenamefont {Khajavikhan}, \citenamefont {Christodoulides},\ and\ \citenamefont {El-Ganainy}}]{zhong2018winding}%
  \BibitemOpen
  \bibfield  {author} {\bibinfo {author} {\bibfnamefont {Q.}~\bibnamefont {Zhong}}, \bibinfo {author} {\bibfnamefont {M.}~\bibnamefont {Khajavikhan}}, \bibinfo {author} {\bibfnamefont {D.~N.}\ \bibnamefont {Christodoulides}},\ and\ \bibinfo {author} {\bibfnamefont {R.}~\bibnamefont {El-Ganainy}},\ }\href {https://doi.org/10.1038/s41467-018-07105-0} {\bibfield  {journal} {\bibinfo  {journal} {Nature Communications}\ }\textbf {\bibinfo {volume} {9}},\ \bibinfo {pages} {4808} (\bibinfo {year} {2018})}\BibitemShut {NoStop}%
\bibitem [{\citenamefont {Ryu}\ \emph {et~al.}(2022)\citenamefont {Ryu}, \citenamefont {Han},\ and\ \citenamefont {Yi}}]{Ryu2022}%
  \BibitemOpen
  \bibfield  {author} {\bibinfo {author} {\bibfnamefont {J.-W.}\ \bibnamefont {Ryu}}, \bibinfo {author} {\bibfnamefont {J.-H.}\ \bibnamefont {Han}},\ and\ \bibinfo {author} {\bibfnamefont {C.-H.}\ \bibnamefont {Yi}},\ }\href {https://doi.org/10.1103/PhysRevA.106.012218} {\bibfield  {journal} {\bibinfo  {journal} {Phys. Rev. A}\ }\textbf {\bibinfo {volume} {106}},\ \bibinfo {pages} {012218} (\bibinfo {year} {2022})}\BibitemShut {NoStop}%
\bibitem [{\citenamefont {Jezequel}\ and\ \citenamefont {Delplace}(2023)}]{Jezequel2023}%
  \BibitemOpen
  \bibfield  {author} {\bibinfo {author} {\bibfnamefont {L.}~\bibnamefont {Jezequel}}\ and\ \bibinfo {author} {\bibfnamefont {P.}~\bibnamefont {Delplace}},\ }\href {https://doi.org/10.1103/PhysRevLett.130.066601} {\bibfield  {journal} {\bibinfo  {journal} {Phys. Rev. Lett.}\ }\textbf {\bibinfo {volume} {130}},\ \bibinfo {pages} {066601} (\bibinfo {year} {2023})}\BibitemShut {NoStop}%
\bibitem [{\citenamefont {Leclerc}\ \emph {et~al.}(2024)\citenamefont {Leclerc}, \citenamefont {Jezequel}, \citenamefont {Perez}, \citenamefont {Bhandare}, \citenamefont {Laibe},\ and\ \citenamefont {Delplace}}]{Leclerc2024}%
  \BibitemOpen
  \bibfield  {author} {\bibinfo {author} {\bibfnamefont {A.}~\bibnamefont {Leclerc}}, \bibinfo {author} {\bibfnamefont {L.}~\bibnamefont {Jezequel}}, \bibinfo {author} {\bibfnamefont {N.}~\bibnamefont {Perez}}, \bibinfo {author} {\bibfnamefont {A.}~\bibnamefont {Bhandare}}, \bibinfo {author} {\bibfnamefont {G.}~\bibnamefont {Laibe}},\ and\ \bibinfo {author} {\bibfnamefont {P.}~\bibnamefont {Delplace}},\ }\href {https://doi.org/10.1103/PhysRevResearch.6.L012055} {\bibfield  {journal} {\bibinfo  {journal} {Phys. Rev. Res.}\ }\textbf {\bibinfo {volume} {6}},\ \bibinfo {pages} {L012055} (\bibinfo {year} {2024})}\BibitemShut {NoStop}%
\bibitem [{\citenamefont {Zhang}\ \emph {et~al.}(2025)\citenamefont {Zhang}, \citenamefont {Solodovchenko}, \citenamefont {Fan}, \citenamefont {Limonov}, \citenamefont {Song}, \citenamefont {Kivshar},\ and\ \citenamefont {Bogdanov}}]{Zhang2025}%
  \BibitemOpen
  \bibfield  {author} {\bibinfo {author} {\bibfnamefont {F.}~\bibnamefont {Zhang}}, \bibinfo {author} {\bibfnamefont {N.~S.}\ \bibnamefont {Solodovchenko}}, \bibinfo {author} {\bibfnamefont {H.}~\bibnamefont {Fan}}, \bibinfo {author} {\bibfnamefont {M.~F.}\ \bibnamefont {Limonov}}, \bibinfo {author} {\bibfnamefont {M.}~\bibnamefont {Song}}, \bibinfo {author} {\bibfnamefont {Y.~S.}\ \bibnamefont {Kivshar}},\ and\ \bibinfo {author} {\bibfnamefont {A.~A.}\ \bibnamefont {Bogdanov}},\ }\href {https://doi.org/10.1126/sciadv.adr9183} {\bibfield  {journal} {\bibinfo  {journal} {Science Advances}\ }\textbf {\bibinfo {volume} {11}},\ \bibinfo {pages} {eadr9183} (\bibinfo {year} {2025})}\BibitemShut {NoStop}%
\bibitem [{\citenamefont {Wiersig}(2014)}]{Wiersig2014}%
  \BibitemOpen
  \bibfield  {author} {\bibinfo {author} {\bibfnamefont {J.}~\bibnamefont {Wiersig}},\ }\href {https://doi.org/10.1103/PhysRevLett.112.203901} {\bibfield  {journal} {\bibinfo  {journal} {Phys. Rev. Lett.}\ }\textbf {\bibinfo {volume} {112}},\ \bibinfo {pages} {203901} (\bibinfo {year} {2014})}\BibitemShut {NoStop}%
\bibitem [{\citenamefont {Chen}\ \emph {et~al.}(2017)\citenamefont {Chen}, \citenamefont {Kaya~Özdemir}, \citenamefont {Zhao}, \citenamefont {Wiersig},\ and\ \citenamefont {Yang}}]{chen2017exceptional}%
  \BibitemOpen
  \bibfield  {author} {\bibinfo {author} {\bibfnamefont {W.}~\bibnamefont {Chen}}, \bibinfo {author} {\bibfnamefont {S.}~\bibnamefont {Kaya~Özdemir}}, \bibinfo {author} {\bibfnamefont {G.}~\bibnamefont {Zhao}}, \bibinfo {author} {\bibfnamefont {J.}~\bibnamefont {Wiersig}},\ and\ \bibinfo {author} {\bibfnamefont {L.}~\bibnamefont {Yang}},\ }\href {https://doi.org/10.1038/nature23281} {\bibfield  {journal} {\bibinfo  {journal} {Nature}\ }\textbf {\bibinfo {volume} {548}},\ \bibinfo {pages} {192} (\bibinfo {year} {2017})}\BibitemShut {NoStop}%
\bibitem [{\citenamefont {Hodaei}\ \emph {et~al.}(2017)\citenamefont {Hodaei}, \citenamefont {Hassan}, \citenamefont {Wittek}, \citenamefont {Garcia-Gracia}, \citenamefont {El-Ganainy}, \citenamefont {Christodoulides},\ and\ \citenamefont {Khajavikhan}}]{hodaei2017enhanced}%
  \BibitemOpen
  \bibfield  {author} {\bibinfo {author} {\bibfnamefont {H.}~\bibnamefont {Hodaei}}, \bibinfo {author} {\bibfnamefont {A.~U.}\ \bibnamefont {Hassan}}, \bibinfo {author} {\bibfnamefont {S.}~\bibnamefont {Wittek}}, \bibinfo {author} {\bibfnamefont {H.}~\bibnamefont {Garcia-Gracia}}, \bibinfo {author} {\bibfnamefont {R.}~\bibnamefont {El-Ganainy}}, \bibinfo {author} {\bibfnamefont {D.~N.}\ \bibnamefont {Christodoulides}},\ and\ \bibinfo {author} {\bibfnamefont {M.}~\bibnamefont {Khajavikhan}},\ }\href {https://doi.org/10.1038/nature23280} {\bibfield  {journal} {\bibinfo  {journal} {Nature}\ }\textbf {\bibinfo {volume} {548}},\ \bibinfo {pages} {187} (\bibinfo {year} {2017})}\BibitemShut {NoStop}%
\bibitem [{\citenamefont {Chen}\ \emph {et~al.}(2019)\citenamefont {Chen}, \citenamefont {Jin},\ and\ \citenamefont {Liu}}]{chen2019sensitivity}%
  \BibitemOpen
  \bibfield  {author} {\bibinfo {author} {\bibfnamefont {C.}~\bibnamefont {Chen}}, \bibinfo {author} {\bibfnamefont {L.}~\bibnamefont {Jin}},\ and\ \bibinfo {author} {\bibfnamefont {R.-B.}\ \bibnamefont {Liu}},\ }\href {https://doi.org/10.1088/1367-2630/ab32ab} {\bibfield  {journal} {\bibinfo  {journal} {New Journal of Physics}\ }\textbf {\bibinfo {volume} {21}},\ \bibinfo {pages} {083002} (\bibinfo {year} {2019})}\BibitemShut {NoStop}%
\bibitem [{\citenamefont {Kononchuk}\ \emph {et~al.}(2022)\citenamefont {Kononchuk}, \citenamefont {Cai}, \citenamefont {Ellis}, \citenamefont {Thevamaran},\ and\ \citenamefont {Kottos}}]{kononchuk2022exceptional}%
  \BibitemOpen
  \bibfield  {author} {\bibinfo {author} {\bibfnamefont {R.}~\bibnamefont {Kononchuk}}, \bibinfo {author} {\bibfnamefont {J.}~\bibnamefont {Cai}}, \bibinfo {author} {\bibfnamefont {F.}~\bibnamefont {Ellis}}, \bibinfo {author} {\bibfnamefont {R.}~\bibnamefont {Thevamaran}},\ and\ \bibinfo {author} {\bibfnamefont {T.}~\bibnamefont {Kottos}},\ }\href {https://doi.org/10.1038/s41586-022-04904-w} {\bibfield  {journal} {\bibinfo  {journal} {Nature}\ }\textbf {\bibinfo {volume} {607}},\ \bibinfo {pages} {697} (\bibinfo {year} {2022})}\BibitemShut {NoStop}%
\bibitem [{\citenamefont {Wiersig}(2020)}]{wiersig2020prospects}%
  \BibitemOpen
  \bibfield  {author} {\bibinfo {author} {\bibfnamefont {J.}~\bibnamefont {Wiersig}},\ }\href {https://doi.org/10.1038/s41467-020-16373-8} {\bibfield  {journal} {\bibinfo  {journal} {Nature Communications}\ }\textbf {\bibinfo {volume} {11}},\ \bibinfo {pages} {2454} (\bibinfo {year} {2020})}\BibitemShut {NoStop}%
\bibitem [{\citenamefont {Longhi}(2015)}]{Longhi_2015}%
  \BibitemOpen
  \bibfield  {author} {\bibinfo {author} {\bibfnamefont {S.}~\bibnamefont {Longhi}},\ }\href {https://doi.org/10.1364/ol.40.005694} {\bibfield  {journal} {\bibinfo  {journal} {Optics Letters}\ }\textbf {\bibinfo {volume} {40}},\ \bibinfo {pages} {5694} (\bibinfo {year} {2015})}\BibitemShut {NoStop}%
\bibitem [{\citenamefont {Brandst\"otter}\ \emph {et~al.}(2019)\citenamefont {Brandst\"otter}, \citenamefont {Makris},\ and\ \citenamefont {Rotter}}]{Brandstotter2019}%
  \BibitemOpen
  \bibfield  {author} {\bibinfo {author} {\bibfnamefont {A.}~\bibnamefont {Brandst\"otter}}, \bibinfo {author} {\bibfnamefont {K.~G.}\ \bibnamefont {Makris}},\ and\ \bibinfo {author} {\bibfnamefont {S.}~\bibnamefont {Rotter}},\ }\href {https://doi.org/10.1103/PhysRevB.99.115402} {\bibfield  {journal} {\bibinfo  {journal} {Phys. Rev. B}\ }\textbf {\bibinfo {volume} {99}},\ \bibinfo {pages} {115402} (\bibinfo {year} {2019})}\BibitemShut {NoStop}%
\bibitem [{\citenamefont {Longhi}\ and\ \citenamefont {Pinotti}(2022)}]{Longhi_2022}%
  \BibitemOpen
  \bibfield  {author} {\bibinfo {author} {\bibfnamefont {S.}~\bibnamefont {Longhi}}\ and\ \bibinfo {author} {\bibfnamefont {E.}~\bibnamefont {Pinotti}},\ }\href {https://doi.org/10.1103/PhysRevB.106.094205} {\bibfield  {journal} {\bibinfo  {journal} {Phys. Rev. B}\ }\textbf {\bibinfo {volume} {106}},\ \bibinfo {pages} {094205} (\bibinfo {year} {2022})}\BibitemShut {NoStop}%
\bibitem [{\citenamefont {Ashida}\ \emph {et~al.}(2020)\citenamefont {Ashida}, \citenamefont {Gong},\ and\ \citenamefont {Ueda}}]{NonHermiPhysics}%
  \BibitemOpen
  \bibfield  {author} {\bibinfo {author} {\bibfnamefont {Y.}~\bibnamefont {Ashida}}, \bibinfo {author} {\bibfnamefont {Z.}~\bibnamefont {Gong}},\ and\ \bibinfo {author} {\bibfnamefont {M.}~\bibnamefont {Ueda}},\ }\href {https://doi.org/10.1080/00018732.2021.1876991} {\bibfield  {journal} {\bibinfo  {journal} {Advances in Physics}\ }\textbf {\bibinfo {volume} {69}},\ \bibinfo {pages} {249–435} (\bibinfo {year} {2020})}\BibitemShut {NoStop}%
\bibitem [{\citenamefont {Yao}\ \emph {et~al.}(2018)\citenamefont {Yao}, \citenamefont {Song},\ and\ \citenamefont {Wang}}]{NonHemChernBands}%
  \BibitemOpen
  \bibfield  {author} {\bibinfo {author} {\bibfnamefont {S.}~\bibnamefont {Yao}}, \bibinfo {author} {\bibfnamefont {F.}~\bibnamefont {Song}},\ and\ \bibinfo {author} {\bibfnamefont {Z.}~\bibnamefont {Wang}},\ }\href {https://doi.org/10.1103/PhysRevLett.121.136802} {\bibfield  {journal} {\bibinfo  {journal} {Phys. Rev. Lett.}\ }\textbf {\bibinfo {volume} {121}},\ \bibinfo {pages} {136802} (\bibinfo {year} {2018})}\BibitemShut {NoStop}%
\bibitem [{\citenamefont {Gong}\ \emph {et~al.}(2018)\citenamefont {Gong}, \citenamefont {Ashida}, \citenamefont {Kawabata}, \citenamefont {Takasan}, \citenamefont {Higashikawa},\ and\ \citenamefont {Ueda}}]{TopoPhase}%
  \BibitemOpen
  \bibfield  {author} {\bibinfo {author} {\bibfnamefont {Z.}~\bibnamefont {Gong}}, \bibinfo {author} {\bibfnamefont {Y.}~\bibnamefont {Ashida}}, \bibinfo {author} {\bibfnamefont {K.}~\bibnamefont {Kawabata}}, \bibinfo {author} {\bibfnamefont {K.}~\bibnamefont {Takasan}}, \bibinfo {author} {\bibfnamefont {S.}~\bibnamefont {Higashikawa}},\ and\ \bibinfo {author} {\bibfnamefont {M.}~\bibnamefont {Ueda}},\ }\href {https://doi.org/10.1103/PhysRevX.8.031079} {\bibfield  {journal} {\bibinfo  {journal} {Phys. Rev. X}\ }\textbf {\bibinfo {volume} {8}},\ \bibinfo {pages} {031079} (\bibinfo {year} {2018})}\BibitemShut {NoStop}%
\bibitem [{\citenamefont {Shen}\ \emph {et~al.}(2018)\citenamefont {Shen}, \citenamefont {Zhen},\ and\ \citenamefont {Fu}}]{TopoBandNonHerm}%
  \BibitemOpen
  \bibfield  {author} {\bibinfo {author} {\bibfnamefont {H.}~\bibnamefont {Shen}}, \bibinfo {author} {\bibfnamefont {B.}~\bibnamefont {Zhen}},\ and\ \bibinfo {author} {\bibfnamefont {L.}~\bibnamefont {Fu}},\ }\href {https://doi.org/10.1103/PhysRevLett.120.146402} {\bibfield  {journal} {\bibinfo  {journal} {Phys. Rev. Lett.}\ }\textbf {\bibinfo {volume} {120}},\ \bibinfo {pages} {146402} (\bibinfo {year} {2018})}\BibitemShut {NoStop}%
\bibitem [{\citenamefont {Zhou}\ \emph {et~al.}(2018)\citenamefont {Zhou}, \citenamefont {Peng}, \citenamefont {Yoon}, \citenamefont {Hsu}, \citenamefont {Nelson}, \citenamefont {Fu}, \citenamefont {Joannopoulos}, \citenamefont {Soljačić},\ and\ \citenamefont {Zhen}}]{zhou2018observation}%
  \BibitemOpen
  \bibfield  {author} {\bibinfo {author} {\bibfnamefont {H.}~\bibnamefont {Zhou}}, \bibinfo {author} {\bibfnamefont {C.}~\bibnamefont {Peng}}, \bibinfo {author} {\bibfnamefont {Y.}~\bibnamefont {Yoon}}, \bibinfo {author} {\bibfnamefont {C.~W.}\ \bibnamefont {Hsu}}, \bibinfo {author} {\bibfnamefont {K.~A.}\ \bibnamefont {Nelson}}, \bibinfo {author} {\bibfnamefont {L.}~\bibnamefont {Fu}}, \bibinfo {author} {\bibfnamefont {J.~D.}\ \bibnamefont {Joannopoulos}}, \bibinfo {author} {\bibfnamefont {M.}~\bibnamefont {Soljačić}},\ and\ \bibinfo {author} {\bibfnamefont {B.}~\bibnamefont {Zhen}},\ }\href {https://doi.org/10.1126/science.aap9859} {\bibfield  {journal} {\bibinfo  {journal} {Science}\ }\textbf {\bibinfo {volume} {359}},\ \bibinfo {pages} {1009–1012} (\bibinfo {year} {2018})}\BibitemShut {NoStop}%
\bibitem [{\citenamefont {Yao}\ and\ \citenamefont {Wang}(2018)}]{Yao2018}%
  \BibitemOpen
  \bibfield  {author} {\bibinfo {author} {\bibfnamefont {S.}~\bibnamefont {Yao}}\ and\ \bibinfo {author} {\bibfnamefont {Z.}~\bibnamefont {Wang}},\ }\href {https://doi.org/10.1103/PhysRevLett.121.086803} {\bibfield  {journal} {\bibinfo  {journal} {Phys. Rev. Lett.}\ }\textbf {\bibinfo {volume} {121}},\ \bibinfo {pages} {086803} (\bibinfo {year} {2018})}\BibitemShut {NoStop}%
\bibitem [{\citenamefont {Okuma}\ \emph {et~al.}(2020)\citenamefont {Okuma}, \citenamefont {Kawabata}, \citenamefont {Shiozaki},\ and\ \citenamefont {Sato}}]{okuma2020topological}%
  \BibitemOpen
  \bibfield  {author} {\bibinfo {author} {\bibfnamefont {N.}~\bibnamefont {Okuma}}, \bibinfo {author} {\bibfnamefont {K.}~\bibnamefont {Kawabata}}, \bibinfo {author} {\bibfnamefont {K.}~\bibnamefont {Shiozaki}},\ and\ \bibinfo {author} {\bibfnamefont {M.}~\bibnamefont {Sato}},\ }\href {https://doi.org/10.1103/PhysRevLett.124.086801} {\bibfield  {journal} {\bibinfo  {journal} {Phys. Rev. Lett.}\ }\textbf {\bibinfo {volume} {124}},\ \bibinfo {pages} {086801} (\bibinfo {year} {2020})}\BibitemShut {NoStop}%
\bibitem [{\citenamefont {Borgnia}\ \emph {et~al.}(2020)\citenamefont {Borgnia}, \citenamefont {Kruchkov},\ and\ \citenamefont {Slager}}]{Borgnia2020}%
  \BibitemOpen
  \bibfield  {author} {\bibinfo {author} {\bibfnamefont {D.~S.}\ \bibnamefont {Borgnia}}, \bibinfo {author} {\bibfnamefont {A.~J.}\ \bibnamefont {Kruchkov}},\ and\ \bibinfo {author} {\bibfnamefont {R.-J.}\ \bibnamefont {Slager}},\ }\href {https://doi.org/10.1103/PhysRevLett.124.056802} {\bibfield  {journal} {\bibinfo  {journal} {Phys. Rev. Lett.}\ }\textbf {\bibinfo {volume} {124}},\ \bibinfo {pages} {056802} (\bibinfo {year} {2020})}\BibitemShut {NoStop}%
\bibitem [{\citenamefont {Zhang}\ \emph {et~al.}(2022{\natexlab{a}})\citenamefont {Zhang}, \citenamefont {Zhang}, \citenamefont {Lu},\ and\ \citenamefont {Chen}}]{zhang2022review}%
  \BibitemOpen
  \bibfield  {author} {\bibinfo {author} {\bibfnamefont {X.}~\bibnamefont {Zhang}}, \bibinfo {author} {\bibfnamefont {T.}~\bibnamefont {Zhang}}, \bibinfo {author} {\bibfnamefont {M.-H.}\ \bibnamefont {Lu}},\ and\ \bibinfo {author} {\bibfnamefont {Y.-F.}\ \bibnamefont {Chen}},\ }\href {https://doi.org/10.1080/23746149.2022.2109431} {\bibfield  {journal} {\bibinfo  {journal} {Advances in Physics: X}\ }\textbf {\bibinfo {volume} {7}},\ \bibinfo {pages} {2109431} (\bibinfo {year} {2022}{\natexlab{a}})}\BibitemShut {NoStop}%
\bibitem [{\citenamefont {Zhang}\ \emph {et~al.}(2022{\natexlab{b}})\citenamefont {Zhang}, \citenamefont {Yang},\ and\ \citenamefont {Fang}}]{zhang2022universal}%
  \BibitemOpen
  \bibfield  {author} {\bibinfo {author} {\bibfnamefont {K.}~\bibnamefont {Zhang}}, \bibinfo {author} {\bibfnamefont {Z.}~\bibnamefont {Yang}},\ and\ \bibinfo {author} {\bibfnamefont {C.}~\bibnamefont {Fang}},\ }\href {https://doi.org/10.1038/s41467-022-30161-6} {\bibfield  {journal} {\bibinfo  {journal} {Nature Communications}\ }\textbf {\bibinfo {volume} {13}},\ \bibinfo {pages} {2496} (\bibinfo {year} {2022}{\natexlab{b}})}\BibitemShut {NoStop}%
\bibitem [{\citenamefont {Wang}\ \emph {et~al.}(2024)\citenamefont {Wang}, \citenamefont {Song},\ and\ \citenamefont {Wang}}]{Wang2024}%
  \BibitemOpen
  \bibfield  {author} {\bibinfo {author} {\bibfnamefont {H.-Y.}\ \bibnamefont {Wang}}, \bibinfo {author} {\bibfnamefont {F.}~\bibnamefont {Song}},\ and\ \bibinfo {author} {\bibfnamefont {Z.}~\bibnamefont {Wang}},\ }\href {https://doi.org/10.1103/PhysRevX.14.021011} {\bibfield  {journal} {\bibinfo  {journal} {Phys. Rev. X}\ }\textbf {\bibinfo {volume} {14}},\ \bibinfo {pages} {021011} (\bibinfo {year} {2024})}\BibitemShut {NoStop}%
\bibitem [{\citenamefont {Hu}(2025)}]{hu2025topological}%
  \BibitemOpen
  \bibfield  {author} {\bibinfo {author} {\bibfnamefont {H.}~\bibnamefont {Hu}},\ }\href {https://doi.org/https://doi.org/10.1016/j.scib.2024.07.022} {\bibfield  {journal} {\bibinfo  {journal} {Science Bulletin}\ }\textbf {\bibinfo {volume} {70}},\ \bibinfo {pages} {51} (\bibinfo {year} {2025})}\BibitemShut {NoStop}%
\bibitem [{\citenamefont {Yang}\ and\ \citenamefont {Bergholtz}(2025)}]{Yang_2025}%
  \BibitemOpen
  \bibfield  {author} {\bibinfo {author} {\bibfnamefont {F.}~\bibnamefont {Yang}}\ and\ \bibinfo {author} {\bibfnamefont {E.~J.}\ \bibnamefont {Bergholtz}},\ }\href {https://doi.org/10.1103/PhysRevResearch.7.023233} {\bibfield  {journal} {\bibinfo  {journal} {Phys. Rev. Res.}\ }\textbf {\bibinfo {volume} {7}},\ \bibinfo {pages} {023233} (\bibinfo {year} {2025})}\BibitemShut {NoStop}%
\bibitem [{\citenamefont {Shen}\ and\ \citenamefont {Lee}(2022)}]{Shen_2022}%
  \BibitemOpen
  \bibfield  {author} {\bibinfo {author} {\bibfnamefont {R.}~\bibnamefont {Shen}}\ and\ \bibinfo {author} {\bibfnamefont {C.~H.}\ \bibnamefont {Lee}},\ }\href {https://doi.org/10.1038/s42005-022-01015-w} {\bibfield  {journal} {\bibinfo  {journal} {Communications Physics}\ }\textbf {\bibinfo {volume} {5}},\ \bibinfo {pages} {238} (\bibinfo {year} {2022})}\BibitemShut {NoStop}%
\bibitem [{\citenamefont {Yang}\ \emph {et~al.}(2022)\citenamefont {Yang}, \citenamefont {Jiang},\ and\ \citenamefont {Bergholtz}}]{Yang_2022}%
  \BibitemOpen
  \bibfield  {author} {\bibinfo {author} {\bibfnamefont {F.}~\bibnamefont {Yang}}, \bibinfo {author} {\bibfnamefont {Q.-D.}\ \bibnamefont {Jiang}},\ and\ \bibinfo {author} {\bibfnamefont {E.~J.}\ \bibnamefont {Bergholtz}},\ }\href {https://doi.org/10.1103/PhysRevResearch.4.023160} {\bibfield  {journal} {\bibinfo  {journal} {Phys. Rev. Res.}\ }\textbf {\bibinfo {volume} {4}},\ \bibinfo {pages} {023160} (\bibinfo {year} {2022})}\BibitemShut {NoStop}%
\bibitem [{\citenamefont {Alsallom}\ \emph {et~al.}(2022)\citenamefont {Alsallom}, \citenamefont {Herviou}, \citenamefont {Yazyev},\ and\ \citenamefont {Brzezi\ifmmode~\acute{n}\else \'{n}\fi{}ska}}]{Alsallom2022}%
  \BibitemOpen
  \bibfield  {author} {\bibinfo {author} {\bibfnamefont {F.}~\bibnamefont {Alsallom}}, \bibinfo {author} {\bibfnamefont {L.}~\bibnamefont {Herviou}}, \bibinfo {author} {\bibfnamefont {O.~V.}\ \bibnamefont {Yazyev}},\ and\ \bibinfo {author} {\bibfnamefont {M.}~\bibnamefont {Brzezi\ifmmode~\acute{n}\else \'{n}\fi{}ska}},\ }\href {https://doi.org/10.1103/PhysRevResearch.4.033122} {\bibfield  {journal} {\bibinfo  {journal} {Phys. Rev. Res.}\ }\textbf {\bibinfo {volume} {4}},\ \bibinfo {pages} {033122} (\bibinfo {year} {2022})}\BibitemShut {NoStop}%
\bibitem [{\citenamefont {Lee}(2016)}]{Lee_2016}%
  \BibitemOpen
  \bibfield  {author} {\bibinfo {author} {\bibfnamefont {T.~E.}\ \bibnamefont {Lee}},\ }\href {https://doi.org/10.1103/PhysRevLett.116.133903} {\bibfield  {journal} {\bibinfo  {journal} {Phys. Rev. Lett.}\ }\textbf {\bibinfo {volume} {116}},\ \bibinfo {pages} {133903} (\bibinfo {year} {2016})}\BibitemShut {NoStop}%
\bibitem [{\citenamefont {Xiong}(2018)}]{xiong2018does}%
  \BibitemOpen
  \bibfield  {author} {\bibinfo {author} {\bibfnamefont {Y.}~\bibnamefont {Xiong}},\ }\href {https://doi.org/10.1088/2399-6528/aab64a} {\bibfield  {journal} {\bibinfo  {journal} {Journal of Physics Communications}\ }\textbf {\bibinfo {volume} {2}},\ \bibinfo {pages} {035043} (\bibinfo {year} {2018})}\BibitemShut {NoStop}%
\bibitem [{\citenamefont {Spring}\ \emph {et~al.}(2024)\citenamefont {Spring}, \citenamefont {Könye}, \citenamefont {Akhmerov},\ and\ \citenamefont {Fulga}}]{Fulga2024}%
  \BibitemOpen
  \bibfield  {author} {\bibinfo {author} {\bibfnamefont {H.}~\bibnamefont {Spring}}, \bibinfo {author} {\bibfnamefont {V.}~\bibnamefont {Könye}}, \bibinfo {author} {\bibfnamefont {A.~R.}\ \bibnamefont {Akhmerov}},\ and\ \bibinfo {author} {\bibfnamefont {I.~C.}\ \bibnamefont {Fulga}},\ }\href {https://doi.org/10.21468/SciPostPhys.17.6.153} {\bibfield  {journal} {\bibinfo  {journal} {SciPost Phys.}\ }\textbf {\bibinfo {volume} {17}},\ \bibinfo {pages} {153} (\bibinfo {year} {2024})}\BibitemShut {NoStop}%
\bibitem [{\citenamefont {Song}\ \emph {et~al.}(2019)\citenamefont {Song}, \citenamefont {Yao},\ and\ \citenamefont {Wang}}]{Song_2019}%
  \BibitemOpen
  \bibfield  {author} {\bibinfo {author} {\bibfnamefont {F.}~\bibnamefont {Song}}, \bibinfo {author} {\bibfnamefont {S.}~\bibnamefont {Yao}},\ and\ \bibinfo {author} {\bibfnamefont {Z.}~\bibnamefont {Wang}},\ }\href {https://doi.org/10.1103/PhysRevLett.123.246801} {\bibfield  {journal} {\bibinfo  {journal} {Phys. Rev. Lett.}\ }\textbf {\bibinfo {volume} {123}},\ \bibinfo {pages} {246801} (\bibinfo {year} {2019})}\BibitemShut {NoStop}%
\bibitem [{\citenamefont {Yang}\ \emph {et~al.}(2020{\natexlab{a}})\citenamefont {Yang}, \citenamefont {Zhang}, \citenamefont {Fang},\ and\ \citenamefont {Hu}}]{Yang2020}%
  \BibitemOpen
  \bibfield  {author} {\bibinfo {author} {\bibfnamefont {Z.}~\bibnamefont {Yang}}, \bibinfo {author} {\bibfnamefont {K.}~\bibnamefont {Zhang}}, \bibinfo {author} {\bibfnamefont {C.}~\bibnamefont {Fang}},\ and\ \bibinfo {author} {\bibfnamefont {J.}~\bibnamefont {Hu}},\ }\href {https://doi.org/10.1103/PhysRevLett.125.226402} {\bibfield  {journal} {\bibinfo  {journal} {Phys. Rev. Lett.}\ }\textbf {\bibinfo {volume} {125}},\ \bibinfo {pages} {226402} (\bibinfo {year} {2020}{\natexlab{a}})}\BibitemShut {NoStop}%
\bibitem [{\citenamefont {Kunst}\ \emph {et~al.}(2018)\citenamefont {Kunst}, \citenamefont {Edvardsson}, \citenamefont {Budich},\ and\ \citenamefont {Bergholtz}}]{Kunst18}%
  \BibitemOpen
  \bibfield  {author} {\bibinfo {author} {\bibfnamefont {F.~K.}\ \bibnamefont {Kunst}}, \bibinfo {author} {\bibfnamefont {E.}~\bibnamefont {Edvardsson}}, \bibinfo {author} {\bibfnamefont {J.~C.}\ \bibnamefont {Budich}},\ and\ \bibinfo {author} {\bibfnamefont {E.~J.}\ \bibnamefont {Bergholtz}},\ }\href {https://doi.org/10.1103/PhysRevLett.121.026808} {\bibfield  {journal} {\bibinfo  {journal} {Phys. Rev. Lett.}\ }\textbf {\bibinfo {volume} {121}},\ \bibinfo {pages} {026808} (\bibinfo {year} {2018})}\BibitemShut {NoStop}%
\bibitem [{\citenamefont {Herviou}\ \emph {et~al.}(2019)\citenamefont {Herviou}, \citenamefont {Bardarson},\ and\ \citenamefont {Regnault}}]{Herviou_2019}%
  \BibitemOpen
  \bibfield  {author} {\bibinfo {author} {\bibfnamefont {L.}~\bibnamefont {Herviou}}, \bibinfo {author} {\bibfnamefont {J.~H.}\ \bibnamefont {Bardarson}},\ and\ \bibinfo {author} {\bibfnamefont {N.}~\bibnamefont {Regnault}},\ }\href {https://doi.org/10.1103/PhysRevA.99.052118} {\bibfield  {journal} {\bibinfo  {journal} {Phys. Rev. A}\ }\textbf {\bibinfo {volume} {99}},\ \bibinfo {pages} {052118} (\bibinfo {year} {2019})}\BibitemShut {NoStop}%
\bibitem [{\citenamefont {Porras}\ and\ \citenamefont {Fern\'andez-Lorenzo}(2019)}]{Porras2019}%
  \BibitemOpen
  \bibfield  {author} {\bibinfo {author} {\bibfnamefont {D.}~\bibnamefont {Porras}}\ and\ \bibinfo {author} {\bibfnamefont {S.}~\bibnamefont {Fern\'andez-Lorenzo}},\ }\href {https://doi.org/10.1103/PhysRevLett.122.143901} {\bibfield  {journal} {\bibinfo  {journal} {Phys. Rev. Lett.}\ }\textbf {\bibinfo {volume} {122}},\ \bibinfo {pages} {143901} (\bibinfo {year} {2019})}\BibitemShut {NoStop}%
\bibitem [{\citenamefont {Brunelli}\ \emph {et~al.}(2023)\citenamefont {Brunelli}, \citenamefont {Wanjura},\ and\ \citenamefont {Nunnenkamp}}]{Brunelli2023}%
  \BibitemOpen
  \bibfield  {author} {\bibinfo {author} {\bibfnamefont {M.}~\bibnamefont {Brunelli}}, \bibinfo {author} {\bibfnamefont {C.~C.}\ \bibnamefont {Wanjura}},\ and\ \bibinfo {author} {\bibfnamefont {A.}~\bibnamefont {Nunnenkamp}},\ }\href {https://doi.org/10.21468/SciPostPhys.15.4.173} {\bibfield  {journal} {\bibinfo  {journal} {SciPost Phys.}\ }\textbf {\bibinfo {volume} {15}},\ \bibinfo {pages} {173} (\bibinfo {year} {2023})}\BibitemShut {NoStop}%
\bibitem [{\citenamefont {Monkman}\ and\ \citenamefont {Sirker}(2025)}]{Monkman_2025}%
  \BibitemOpen
  \bibfield  {author} {\bibinfo {author} {\bibfnamefont {K.}~\bibnamefont {Monkman}}\ and\ \bibinfo {author} {\bibfnamefont {J.}~\bibnamefont {Sirker}},\ }\href {https://doi.org/10.1103/PhysRevLett.134.056601} {\bibfield  {journal} {\bibinfo  {journal} {Phys. Rev. Lett.}\ }\textbf {\bibinfo {volume} {134}},\ \bibinfo {pages} {056601} (\bibinfo {year} {2025})}\BibitemShut {NoStop}%
\bibitem [{\citenamefont {Mardani}\ \emph {et~al.}()\citenamefont {Mardani}, \citenamefont {Pimenta},\ and\ \citenamefont {Sirker}}]{mardani2025}%
  \BibitemOpen
  \bibfield  {author} {\bibinfo {author} {\bibfnamefont {Y.}~\bibnamefont {Mardani}}, \bibinfo {author} {\bibfnamefont {R.~A.}\ \bibnamefont {Pimenta}},\ and\ \bibinfo {author} {\bibfnamefont {J.}~\bibnamefont {Sirker}},\ }\href {https://arxiv.org/abs/2410.01542} {}\Eprint {https://arxiv.org/abs/2410.01542} {arXiv:2410.01542} \BibitemShut {NoStop}%
\bibitem [{\citenamefont {Trefethen}\ and\ \citenamefont {Embree}(2020)}]{trefethen2020spectra}%
  \BibitemOpen
  \bibfield  {author} {\bibinfo {author} {\bibfnamefont {L.~N.}\ \bibnamefont {Trefethen}}\ and\ \bibinfo {author} {\bibfnamefont {M.}~\bibnamefont {Embree}},\ }\href@noop {} {\emph {\bibinfo {title} {Spectra and pseudospectra: the behavior of nonnormal matrices and operators}}}\ (\bibinfo  {publisher} {Princeton university press},\ \bibinfo {year} {2020})\BibitemShut {NoStop}%
\bibitem [{\citenamefont {Flynn}\ \emph {et~al.}(2021)\citenamefont {Flynn}, \citenamefont {Cobanera},\ and\ \citenamefont {Viola}}]{flynn2023topological}%
  \BibitemOpen
  \bibfield  {author} {\bibinfo {author} {\bibfnamefont {V.~P.}\ \bibnamefont {Flynn}}, \bibinfo {author} {\bibfnamefont {E.}~\bibnamefont {Cobanera}},\ and\ \bibinfo {author} {\bibfnamefont {L.}~\bibnamefont {Viola}},\ }\href {https://doi.org/10.1103/PhysRevLett.127.245701} {\bibfield  {journal} {\bibinfo  {journal} {Phys. Rev. Lett.}\ }\textbf {\bibinfo {volume} {127}},\ \bibinfo {pages} {245701} (\bibinfo {year} {2021})}\BibitemShut {NoStop}%
\bibitem [{\citenamefont {Okuma}\ and\ \citenamefont {Sato}(2021)}]{Okuma2021}%
  \BibitemOpen
  \bibfield  {author} {\bibinfo {author} {\bibfnamefont {N.}~\bibnamefont {Okuma}}\ and\ \bibinfo {author} {\bibfnamefont {M.}~\bibnamefont {Sato}},\ }\href {https://doi.org/10.1103/PhysRevB.103.085428} {\bibfield  {journal} {\bibinfo  {journal} {Phys. Rev. B}\ }\textbf {\bibinfo {volume} {103}},\ \bibinfo {pages} {085428} (\bibinfo {year} {2021})}\BibitemShut {NoStop}%
\bibitem [{\citenamefont {Hatano}\ and\ \citenamefont {Nelson}(1996)}]{HatanoNelson}%
  \BibitemOpen
  \bibfield  {author} {\bibinfo {author} {\bibfnamefont {N.}~\bibnamefont {Hatano}}\ and\ \bibinfo {author} {\bibfnamefont {D.~R.}\ \bibnamefont {Nelson}},\ }\href {https://doi.org/10.1103/PhysRevLett.77.570} {\bibfield  {journal} {\bibinfo  {journal} {Phys. Rev. Lett.}\ }\textbf {\bibinfo {volume} {77}},\ \bibinfo {pages} {570} (\bibinfo {year} {1996})}\BibitemShut {NoStop}%
\bibitem [{\citenamefont {Hatano}\ and\ \citenamefont {Nelson}(1998)}]{hatano1998non}%
  \BibitemOpen
  \bibfield  {author} {\bibinfo {author} {\bibfnamefont {N.}~\bibnamefont {Hatano}}\ and\ \bibinfo {author} {\bibfnamefont {D.~R.}\ \bibnamefont {Nelson}},\ }\href {https://doi.org/10.1103/PhysRevB.58.8384} {\bibfield  {journal} {\bibinfo  {journal} {Phys. Rev. B}\ }\textbf {\bibinfo {volume} {58}},\ \bibinfo {pages} {8384} (\bibinfo {year} {1998})}\BibitemShut {NoStop}%
\bibitem [{\citenamefont {Feng}\ \emph {et~al.}(2025)\citenamefont {Feng}, \citenamefont {Liu}, \citenamefont {Zhang},\ and\ \citenamefont {Chen}}]{Feng_2025}%
  \BibitemOpen
  \bibfield  {author} {\bibinfo {author} {\bibfnamefont {X.}~\bibnamefont {Feng}}, \bibinfo {author} {\bibfnamefont {S.}~\bibnamefont {Liu}}, \bibinfo {author} {\bibfnamefont {S.-X.}\ \bibnamefont {Zhang}},\ and\ \bibinfo {author} {\bibfnamefont {S.}~\bibnamefont {Chen}},\ }\bibfield  {journal} {\bibinfo  {journal} {Physical Review B}\ }\textbf {\bibinfo {volume} {111}},\ \href {https://doi.org/10.1103/g1cw-tk7f} {10.1103/g1cw-tk7f} (\bibinfo {year} {2025})\BibitemShut {NoStop}%
\bibitem [{\citenamefont {Yang}\ \emph {et~al.}(2020{\natexlab{b}})\citenamefont {Yang}, \citenamefont {Zhang}, \citenamefont {Fang},\ and\ \citenamefont {Hu}}]{PhysRevLett.125.226402}%
  \BibitemOpen
  \bibfield  {author} {\bibinfo {author} {\bibfnamefont {Z.}~\bibnamefont {Yang}}, \bibinfo {author} {\bibfnamefont {K.}~\bibnamefont {Zhang}}, \bibinfo {author} {\bibfnamefont {C.}~\bibnamefont {Fang}},\ and\ \bibinfo {author} {\bibfnamefont {J.}~\bibnamefont {Hu}},\ }\href {https://doi.org/10.1103/PhysRevLett.125.226402} {\bibfield  {journal} {\bibinfo  {journal} {Phys. Rev. Lett.}\ }\textbf {\bibinfo {volume} {125}},\ \bibinfo {pages} {226402} (\bibinfo {year} {2020}{\natexlab{b}})}\BibitemShut {NoStop}%
\bibitem [{\citenamefont {Soares}\ and\ \citenamefont {Schirò}(2024)}]{10.21468/SciPostPhys.17.5.128}%
  \BibitemOpen
  \bibfield  {author} {\bibinfo {author} {\bibfnamefont {R.~D.}\ \bibnamefont {Soares}}\ and\ \bibinfo {author} {\bibfnamefont {M.}~\bibnamefont {Schirò}},\ }\href {https://doi.org/10.21468/SciPostPhys.17.5.128} {\bibfield  {journal} {\bibinfo  {journal} {SciPost Phys.}\ }\textbf {\bibinfo {volume} {17}},\ \bibinfo {pages} {128} (\bibinfo {year} {2024})}\BibitemShut {NoStop}%
\bibitem [{\citenamefont {Liu}\ \emph {et~al.}(2019)\citenamefont {Liu}, \citenamefont {Jiang},\ and\ \citenamefont {Chen}}]{Liu2019}%
  \BibitemOpen
  \bibfield  {author} {\bibinfo {author} {\bibfnamefont {C.-H.}\ \bibnamefont {Liu}}, \bibinfo {author} {\bibfnamefont {H.}~\bibnamefont {Jiang}},\ and\ \bibinfo {author} {\bibfnamefont {S.}~\bibnamefont {Chen}},\ }\href {https://doi.org/10.1103/PhysRevB.99.125103} {\bibfield  {journal} {\bibinfo  {journal} {Phys. Rev. B}\ }\textbf {\bibinfo {volume} {99}},\ \bibinfo {pages} {125103} (\bibinfo {year} {2019})}\BibitemShut {NoStop}%
\bibitem [{\citenamefont {Schomerus}(2022)}]{Henning2022}%
  \BibitemOpen
  \bibfield  {author} {\bibinfo {author} {\bibfnamefont {H.}~\bibnamefont {Schomerus}},\ }\href {https://doi.org/10.1103/PhysRevA.106.063509} {\bibfield  {journal} {\bibinfo  {journal} {Phys. Rev. A}\ }\textbf {\bibinfo {volume} {106}},\ \bibinfo {pages} {063509} (\bibinfo {year} {2022})}\BibitemShut {NoStop}%
\bibitem [{\citenamefont {Kane}\ and\ \citenamefont {Lubensky}(2013)}]{KaneLubensky}%
  \BibitemOpen
  \bibfield  {author} {\bibinfo {author} {\bibfnamefont {C.~L.}\ \bibnamefont {Kane}}\ and\ \bibinfo {author} {\bibfnamefont {T.~C.}\ \bibnamefont {Lubensky}},\ }\href {https://doi.org/10.1038/nphys2835} {\bibfield  {journal} {\bibinfo  {journal} {Nature Physics}\ }\textbf {\bibinfo {volume} {10}},\ \bibinfo {pages} {39–45} (\bibinfo {year} {2013})}\BibitemShut {NoStop}%
\bibitem [{\citenamefont {Jezequel}\ and\ \citenamefont {Delplace}(2024)}]{jezequel2024}%
  \BibitemOpen
  \bibfield  {author} {\bibinfo {author} {\bibfnamefont {L.}~\bibnamefont {Jezequel}}\ and\ \bibinfo {author} {\bibfnamefont {P.}~\bibnamefont {Delplace}},\ }\href {https://doi.org/10.21468/SciPostPhys.17.2.060} {\bibfield  {journal} {\bibinfo  {journal} {SciPost Phys.}\ }\textbf {\bibinfo {volume} {17}},\ \bibinfo {pages} {060} (\bibinfo {year} {2024})}\BibitemShut {NoStop}%
\bibitem [{\citenamefont {Zhang}\ \emph {et~al.}(2020)\citenamefont {Zhang}, \citenamefont {Yang},\ and\ \citenamefont {Fang}}]{Zhang_2020}%
  \BibitemOpen
  \bibfield  {author} {\bibinfo {author} {\bibfnamefont {K.}~\bibnamefont {Zhang}}, \bibinfo {author} {\bibfnamefont {Z.}~\bibnamefont {Yang}},\ and\ \bibinfo {author} {\bibfnamefont {C.}~\bibnamefont {Fang}},\ }\href {https://doi.org/10.1103/PhysRevLett.125.126402} {\bibfield  {journal} {\bibinfo  {journal} {Phys. Rev. Lett.}\ }\textbf {\bibinfo {volume} {125}},\ \bibinfo {pages} {126402} (\bibinfo {year} {2020})}\BibitemShut {NoStop}%
\bibitem [{\citenamefont {Yin}\ \emph {et~al.}(2018)\citenamefont {Yin}, \citenamefont {Jiang}, \citenamefont {Li}, \citenamefont {L\"u},\ and\ \citenamefont {Chen}}]{Yin2018}%
  \BibitemOpen
  \bibfield  {author} {\bibinfo {author} {\bibfnamefont {C.}~\bibnamefont {Yin}}, \bibinfo {author} {\bibfnamefont {H.}~\bibnamefont {Jiang}}, \bibinfo {author} {\bibfnamefont {L.}~\bibnamefont {Li}}, \bibinfo {author} {\bibfnamefont {R.}~\bibnamefont {L\"u}},\ and\ \bibinfo {author} {\bibfnamefont {S.}~\bibnamefont {Chen}},\ }\href {https://doi.org/10.1103/PhysRevA.97.052115} {\bibfield  {journal} {\bibinfo  {journal} {Phys. Rev. A}\ }\textbf {\bibinfo {volume} {97}},\ \bibinfo {pages} {052115} (\bibinfo {year} {2018})}\BibitemShut {NoStop}%
\bibitem [{\citenamefont {Lin}\ \emph {et~al.}(2023)\citenamefont {Lin}, \citenamefont {Tai}, \citenamefont {Li},\ and\ \citenamefont {Lee}}]{lin_topological_2023}%
  \BibitemOpen
  \bibfield  {author} {\bibinfo {author} {\bibfnamefont {R.}~\bibnamefont {Lin}}, \bibinfo {author} {\bibfnamefont {T.}~\bibnamefont {Tai}}, \bibinfo {author} {\bibfnamefont {L.}~\bibnamefont {Li}},\ and\ \bibinfo {author} {\bibfnamefont {C.~H.}\ \bibnamefont {Lee}},\ }\href {https://doi.org/10.1007/s11467-023-1309-z} {\bibfield  {journal} {\bibinfo  {journal} {Frontiers of Physics}\ }\textbf {\bibinfo {volume} {18}},\ \bibinfo {pages} {53605} (\bibinfo {year} {2023})}\BibitemShut {NoStop}%
\bibitem [{\citenamefont {Haga}\ \emph {et~al.}(2021)\citenamefont {Haga}, \citenamefont {Nakagawa}, \citenamefont {Hamazaki},\ and\ \citenamefont {Ueda}}]{PhysRevLett.127.070402}%
  \BibitemOpen
  \bibfield  {author} {\bibinfo {author} {\bibfnamefont {T.}~\bibnamefont {Haga}}, \bibinfo {author} {\bibfnamefont {M.}~\bibnamefont {Nakagawa}}, \bibinfo {author} {\bibfnamefont {R.}~\bibnamefont {Hamazaki}},\ and\ \bibinfo {author} {\bibfnamefont {M.}~\bibnamefont {Ueda}},\ }\href {https://doi.org/10.1103/PhysRevLett.127.070402} {\bibfield  {journal} {\bibinfo  {journal} {Phys. Rev. Lett.}\ }\textbf {\bibinfo {volume} {127}},\ \bibinfo {pages} {070402} (\bibinfo {year} {2021})}\BibitemShut {NoStop}%
\bibitem [{\citenamefont {Zheng}\ \emph {et~al.}(2024)\citenamefont {Zheng}, \citenamefont {Qiao}, \citenamefont {Wang}, \citenamefont {Cao},\ and\ \citenamefont {Chen}}]{Zheng2024}%
  \BibitemOpen
  \bibfield  {author} {\bibinfo {author} {\bibfnamefont {M.}~\bibnamefont {Zheng}}, \bibinfo {author} {\bibfnamefont {Y.}~\bibnamefont {Qiao}}, \bibinfo {author} {\bibfnamefont {Y.}~\bibnamefont {Wang}}, \bibinfo {author} {\bibfnamefont {J.}~\bibnamefont {Cao}},\ and\ \bibinfo {author} {\bibfnamefont {S.}~\bibnamefont {Chen}},\ }\href {https://doi.org/10.1103/PhysRevLett.132.086502} {\bibfield  {journal} {\bibinfo  {journal} {Phys. Rev. Lett.}\ }\textbf {\bibinfo {volume} {132}},\ \bibinfo {pages} {086502} (\bibinfo {year} {2024})}\BibitemShut {NoStop}%
\bibitem [{\citenamefont {Hamanaka}\ and\ \citenamefont {Kawabata}(2025)}]{Hamanaka2025}%
  \BibitemOpen
  \bibfield  {author} {\bibinfo {author} {\bibfnamefont {S.}~\bibnamefont {Hamanaka}}\ and\ \bibinfo {author} {\bibfnamefont {K.}~\bibnamefont {Kawabata}},\ }\href {https://doi.org/10.1103/PhysRevB.111.035144} {\bibfield  {journal} {\bibinfo  {journal} {Phys. Rev. B}\ }\textbf {\bibinfo {volume} {111}},\ \bibinfo {pages} {035144} (\bibinfo {year} {2025})}\BibitemShut {NoStop}%
\bibitem [{\citenamefont {Zhong}\ \emph {et~al.}(2025)\citenamefont {Zhong}, \citenamefont {Pan}, \citenamefont {Lin}, \citenamefont {Wang},\ and\ \citenamefont {Hu}}]{Zhong2025}%
  \BibitemOpen
  \bibfield  {author} {\bibinfo {author} {\bibfnamefont {P.}~\bibnamefont {Zhong}}, \bibinfo {author} {\bibfnamefont {W.}~\bibnamefont {Pan}}, \bibinfo {author} {\bibfnamefont {H.}~\bibnamefont {Lin}}, \bibinfo {author} {\bibfnamefont {X.}~\bibnamefont {Wang}},\ and\ \bibinfo {author} {\bibfnamefont {S.}~\bibnamefont {Hu}},\ }\href {https://doi.org/10.1103/5vnl-w9p4} {\bibfield  {journal} {\bibinfo  {journal} {Phys. Rev. Lett.}\ }\textbf {\bibinfo {volume} {135}},\ \bibinfo {pages} {106502} (\bibinfo {year} {2025})}\BibitemShut {NoStop}%
\end{thebibliography}%
\onecolumngrid
\appendix
\section{Equivalence between the different notions of spectrum}
\label{ap:B}

In this appendix, we demonstrate that the eigenstate approach, the singular value spectrum, and the bounded inverse spectrum discussed in the main text provide equivalent characterizations of a spectral gap. We focus exclusively on the \textit{right gap} case; the proof for the left gap follows identically upon taking $H' = H^\dagger$.

Let $A = H - E$. For $\Delta > 0$, we want to show that the following three criteria are equivalent:
\begin{align}
\text{Pseudospectrum criteria:} \quad
&\forall |\psi\rangle, \; \| A |\psi\rangle \| \geq \Delta \| |\psi\rangle \|\label{eq:B1}.  \\
\text{Singular value criteria:} \quad
&\sigma_{\min}(A^\dagger A) \geq \Delta^2.\label{eq:B2}  \\
\text{Bounded inverse criteria:} \quad
&\forall |\phi\rangle, \; \| A^{-1} |\phi\rangle \| \leq \Delta^{-1} \| |\phi\rangle \|.\label{eq:B3}
\end{align}

\paragraph{Equivalence between \eqref{eq:B1} and \eqref{eq:B2}:}  
Squaring the inequality in \eqref{eq:B1} gives 
\begin{equation}
\|A\psi\|^2 = \langle \psi | A^\dagger A | \psi \rangle \ge \Delta^2 \langle \psi | \psi \rangle \quad \forall |\psi\rangle.\label{eq:B4}
\end{equation}
This inequality means that the operator $A^\dagger A$ has all eigenvalues greater than or equal to $\Delta^2$, which in turn implies that its smallest eigenvalue satisfies $\sigma_{\mathrm{min}}(A^\dagger A) \ge \Delta^2$. Since the smallest singular value squared of $A$ equals the smallest eigenvalue of $A^\dagger A$, that is, $\sigma_{\mathrm{min}}(A^\dagger A)$, we recover \eqref{eq:B2}. Conversely if $\sigma_{\mathrm{min}}(A^\dagger A)\geq\Delta ^2$, we have the inequality \eqref{eq:B4} and therefore \eqref{eq:B1}.

\paragraph{Equivalence between \eqref{eq:B1} and \eqref{eq:B3}:}  
Suppose \eqref{eq:B1} holds. For any state $|\phi\rangle$, define $|\psi\rangle = A^{-1}|\phi\rangle$. Substituting into \eqref{eq:B1} yields
\begin{equation}
\| \phi \| = \| A \psi \| \ge \Delta \| \psi \| = \Delta \| A^{-1} \phi \|,
\end{equation}
which implies $\|A^{-1} \phi\| \le \Delta^{-1} \|\phi\|$, and therefore (B3). Conversely, if \eqref{eq:B3} holds, set $|\phi\rangle = A|\psi\rangle$. Then \eqref{eq:B3} gives $\|\psi\| \le \Delta^{-1} \|A\psi\|$, which is exactly \eqref{eq:B1}.

Since \eqref{eq:B1} is equivalent to both \eqref{eq:B2} and \eqref{eq:B3}, all three criteria are mathematically equivalent.

\section{Eigenstate spectrum and open boundary conditions}
\label{ap:A}
In this appendix we show that for any local non-Hermitian Hamiltonian, the eigenstate spectrum has the same property as the Hermitian spectrum in that the open boundary condition spectrum includes, in the thermodynamic limit, the bulk bands of the periodic boundary condition spectrum
For periodic Hamiltonian $H_\text{PBC}$ we can make the distinction between the unit cell number $n$ and internal degrees of freedom $\alpha$ of the unit cell on which the Hamiltonian is periodic.

The eigenstate of the periodic Hamiltonian will then take the form of Bloch state with 
\begin{equation}
    \ket{\psi} = \sum_x e^{ikx}c_\alpha\ket{x,\alpha}
\end{equation}
with quasimomentum $k$ and energy $E$.

If we then consider the same Hamiltonian but with open-boundary condition $H_\text{OBC}$, we can construct from this Bloch state, a state $\ket{\psi'}$ that is an eigenstate of $H_\text{OBC}$ up to an error that vanishes in the large length limit. This state is the following 
\begin{equation}
\ket{\psi'} = \sum_{x,\alpha} e^{ikx}g(x/L)c_\alpha\ket{x,\alpha}= g(\hat{x}/L)\ket{\psi}
\end{equation}
where $g(x/L)$ is a smooth envelope function vanishing at the boundaries.
The slow variation of $g(x/L)$ on the scale of the lattice constant ensures that $\|[H_\text{OBC},g(\hat{x}/L)]\|%
    =O(1/L).$
This, together with the vanishing of $g(x/L)$ at the boundaries, ensures that the amplitude of $\ket{\psi'}$ is negligible near the boundary, which suppresses boundary-related errors by a factor $O(1/L)$. This implies that the resulting state has a small energy variance: 
\begin{Aligned}
    &\|(H_\text{OBC}-E)g(\hat{x}/L)\ket{\psi}\|=  \|(H_\text{PBC}-E)g(\hat{x}/L)\ket{\psi}\|+O(1/L)\\
    &=\|g(\hat{x}/L)(H_\text{PBC}-E)\ket{\psi}\|+\|[H_\text{PBC},g(\hat{x}/L)]\ket{\psi}\|+ O(1/L)=O(1/L)
\end{Aligned} 
where we used that $(H_\text{PBC}-E)\ket{\psi}=0$ as $\ket{\psi}$ is an eigenstate of the periodic Hamiltonian.
This proves that all periodic bulk eigenstates have open-boundary counterparts in the eigenstate spectrum for large enough systems.

\end{document}